\documentclass[11pt]{report}
\usepackage{latexsym} % Gets \Box etc
\usepackage{amssymb}  % \gtrsim, \geqslant, etc etc: see amsguide.ps
\usepackage{amsbsy}   % \pmb and \boldsymbol
\usepackage{graphicx}       % For PostScript figures
\usepackage{bm}   % Bold math: \bm{1}
\usepackage{mathdots}
\usepackage[title,titletoc,header]{appendix}
\usepackage{color}

\newcommand{\MS}{{\rm MathieuSin}}
\newcommand{\MC}{{\rm MathieuCos}}
\newcommand{\p}{\partial}
\newcommand{\<}{\langle}
\renewcommand{\>}{\rangle} % LaTeX: \> already defined

\newcommand{\nn}{\nonumber \\}
\newcommand{\be}[1]{\begin{equation}\label{#1}}
\newcommand{\ee}{\end{equation}}
\newcommand{\ba}[1]{\begin{eqnarray}\label{#1}}
\newcommand{\ea}{\end{eqnarray}}

\makeatletter %\catcode`\@=11

\begin{document}
\bibliographystyle{unsrt}
%\preprint{}

\title{\bf Speeding Up Classical and Quantum Adiabatic Processes: Implications for Work Functions and Heat Engine Designs\footnote{This is a rough progress report. Manuscripts are being prepared for submission for publication.}}

\author{Jia-wen Deng, Qing-hai Wang, and Jiangbin Gong}

\date{May 16, 2013}

%\begin{titlepage}
\maketitle

\begin{abstract}
  Adiabatic processes are important for studying the dynamics of a time-dependent system. Conventionally, the adiabatic processes can only be achieved by varying the system slowly. We speed up both classical and quantum adiabatic processes by adding control protocols. In classical systems, we work out the control protocols by analyzing the classical adiabatic approximation. In quantum systems, we follow the idea of transitionless driving by Berry [{\it J.~Phys.~A: Math.~Theor.} {\bf 42} 365303 (2009)]. Such fast-forward adiabatic processes can be performed at arbitrary fast speed, and in the meanwhile reduce the work fluctuation. In both systems, we use a time-dependent harmonic oscillator model to work out explicitly the work function and the work fluctuation in three types of processes: fast-forward adiabatic processes, adiabatic processes, and non-adiabatic processes. We show the significant reduction on work fluctuation in fast-forward adiabatic process. We further illustrate how the fast-forward process improved the converging rate of the Jarzynski equality between the work function and the free energy. As an application, we show that the fast-forward process not only maximizes the output power but also improve the efficiency of a quantum engine.
\end{abstract}
\tableofcontents
%\end{titlepage}

% \maketitle

%%%%%%%%%%%%%%%%%%%%%%%%%%%%%%
\chapter{Introduction}
%%%%%%%%%%%%%%%%%%%%%%%%%%%

For all thermodynamical systems, the macroscopic quantities have a fluctuation because of the statistical nature. According to the law of large numbers, the fluctuation is negligible for large system, which means the probability distribution concentrates near the expectation \cite{Huang}. But  for small systems, the macroscopic quantity spreads in a wide range,which urges us to explore more on the distribution of the quantity.

The probability distribution of the work done to the system under a certain process is usually referred as work function. Work function, together with work fluctuation of small system have attracted much attention recently \cite{JarFluc,HanReview}. Work function also relates non-equilibrium qualities with the equilibrium ones \cite{Jar}. For example, Jarzynski equality relates the non-equilibrium work $W$ with Helmholtz free energy $F$ through $\<e^{-\beta W}\>=e^{-\beta \Delta F}$.  In such discussions, the work fluctuation becomes a vital issue because it gives us information about the error in the estimation of $\<e^{-\beta W}\>$ in practice. Therefore understanding the work function \cite{HanReview,HanOpen}, as well as suppressing the corresponding work fluctuation are very important for small systems.

Researchers are making significant progress on work function. Some recent researches \cite{Lutz} compare the work function of adiabatic and non-adiabatic process under quantum scheme. Results show that adiabatic process owns smaller work fluctuation. This result is not surprising, because adiabatic process will keep the population on each state invariant, or in other words, eliminate transitions between the eigenstates of the system.

However, conventional adiabatic process requires the parameter changing slowly, and due to this reason, it will take a comparatively long time period in practice. Thus one of our motivations is to speed up adiabatic process. To be more precise, in quantum case, we hope to eliminate the transition between states even if the parameter changes rapidly. And in classical case, we will keep the action variable, a classical analog of quantum number invariant as time evolves. We notice that in both cases, we are trying to accomplish a transitionless feature. Based on the previous works of transitionless driving \cite{Berry}, we develop a method to achieved this goal  in both quantum and classical cases by adding a control field to the system. With this approach, the system effectively undergoes an adiabatic process in a short time period, which is definitely a powerful tool for practical purpose.

Based on recent works on work function and Jarzynski equality, we digest deeper on this topic, and use an extra driving field to achieve the so-called fast-forward adiabatic process. In the mean time, the fast-forward adiabatic process could retain all the features of the work function and work fluctuation of conventional adiabatic process with a carefully chosen control field. One amazing result is the estimation of $\<e^{-\beta W}\>$ converges much faster in practice with such control field.

Fast-forward adiabatic process also has potential applications in technology aspect. Recent research on quantum Otto engine \cite{LutzEng} is faced with choices between efficiency and output power. In the conventional scheme, non-adiabatic cycles have smaller efficiency but larger output power, compared with adiabatic cycles. Qualitatively, non-adiabatic cycles have larger work fluctuation thus might not be very efficient; but they can be performed within arbitrarily short duration time, thus the output power could be very large. However, if we remember the previously mentioned remarkable features of our fast-forward adiabatic process, we realize that it minimizes the duration time and work fluctuation at the same time. Follow the same logic, in later chapters we could see how our fast-forward adiabatic process helps the quantum engine to achieve the maximum efficiency and output power at the same time.

In the rest of this report, we will first review both quantum and classical adiabatic theorem in the second chapter, followed by the formal definitions and discussions on work function and work fluctuation in the third chapter. After that, we will introduce our original work on classical fast-forward adiabatic process, including the formal solution of control field and application in 1-D harmonic oscillator. Work functions of adiabatic and non-adiabatic processes will be compared in analytical and numerical manner. Next, for the quantum fast-forward adiabatic process, we will follow Berry\rq{}s approach of transitionless driving. Furthermore, we will consider its work function and compare it with quantum non-adiabatic process in a similar way. Last but not least, we will show some dramatic application of our fast-forward adiabatic process, including increasing the converging speed of $\<e^{-\beta W}\>$ and improving the performance of quantum engine.

%%%%%%%%%%%%%%%%%%%%%%%%%%%%%%%%%
\chapter{Adiabatic Theorem}
%%%%%%%%%%%%%%%%%%%%%%%%%%%%%%%%%%%
Adiabatic process plays an important role in modern quantum mechanics. Because of the population-invariant nature of adiabatic process, it is widely used in quantum optics and atomic physics in both theoretical \cite{Chen2}\cite{ChenTran} and experimental aspect \cite{Bouch}. Besides that, there are some very fundamental signatures of a quantum system, for example, Berry\rq{}s phase, can only be described and measured when the system undergoes a cyclic adiabatic process.

Adiabatic theorem points out one way of realizing the adiabatic process. It tells us that a system usually undergoes an adiabatic process when the parameters of the system are changing slowly. Thus slowly changing the parameters becomes the most common approach to adiabatic process. Such approach will be referred as conventional adiabatic process in the rest of this article.

In this chapter, we will review both quantum and classical adiabatic theorem to explain why the changing rate of parameter matters. Particularly for classical adiabatic theorem, before constructing fast-adiabatic process, we hope to introduce an unfamiliar tool called action-angle variables and make analog with the more familiar quantum version.

\section{Quantum Adiabatic Theorem}
We will illustrate the quantum adiabatic theorem for the system with only one time-dependent parameter. Systems with more parameters are quite similar.

For such system, Hamiltonian is represented by $\hat H_0(\lambda(t))$, where $\lambda(t)$ is the time-dependent parameter. Notice here we write it as $\hat H_0$ rather than $\hat H$. This is because in later chapters we will modify $\hat H_0$ and hope the notations to be consistent with each other.

\subsection{Adiabatic Approximation}
The state at time $t$ satisfies
\begin{equation}
{i} \hbar{\partial \over \partial t}| \Psi(t)\> = \hat H_0(\lambda(t)) |\Psi(t)\>,
\label{c1shordinger}
\end{equation}
and instantaneous eigenstates of $\hat H_0(\lambda(t))$ are given by
\begin{equation}
E_n(t)|n(t)\> = \hat H_0(\lambda(t)) |n(t)\>.
\label{c1staticstate}
\end{equation}
The general solution of $|\Psi(t)\>$ can be expanded using eigenstates of $\hat H_0(\lambda(t))$ at time $t$, i.e.,
\begin{eqnarray}
|\Psi(t)\> &=& \sum\limits_{n}  |n(t)\>\<n(t)|\Psi(t)\>\nn
                &\equiv& \sum\limits_{n} C_n(t) |n(t)\> e^{{i}\theta_n(t)},
                \label{c1generalsolution}
\end{eqnarray}
where
\begin{equation}
\theta_n(t) = - {1\over\hbar}\int_0^t E_n(t\rq{})dt\rq{}
\end{equation}
is the dynamical phase of state $|n(t)\>$ at $t$. Plug (\ref{c1generalsolution}) into (\ref{c1shordinger}),
\begin{eqnarray}
{i}\hbar\sum\limits_{n}\left( \dot C_n |n(t)\> +C_n|\partial_tn(t)\>+  {i}C_n|n(t)\>\dot\theta_n \right)e^{{i}\theta_n} &=& \sum\limits_{n}C_n\hat H_0(t) |n(t)\> e^{{i}\theta_n}   \nn
\sum\limits_{n}\left( \dot C_n |n(t)\> +C_n |\partial_tn(t)\>\right)e^{{i}\theta_n} &=&0 \nn
\<m(t)|\sum\limits_{n}\left( \dot C_n |n(t)\> +C_n |\partial_tn(t)\>\right)e^{{i}\theta_n} &=&0 \nn
-\sum\limits_{n}C_n \<m(t)|\partial_t n(t)\>e^{{i}\theta_n}&=&\dot C_m.
\label{c1coefficient}
\end{eqnarray}
Here we omit $t$ in time-dependent terms $C(t)$ and $\theta(t)$ for convenience. Differentiating (\ref{c1staticstate}) on both sides gives us
\begin{equation}
\partial_t E_n(t)|n(t)\>+E_n(t)|\partial_t n(t)\> =  \partial_t \hat H_0(t) |n(t)\>+\hat H_0(t) |\partial_t n(t)\>,
\end{equation}
and multiplying $\<m(t)|$ $(m\neq n)$ on the left then gives
\begin{eqnarray}
E_n \<m|\partial_t n\> &=&  \<m|\partial_t \hat H_0(\lambda(t)) |n\> +E_m\<m|\partial_t n\> \nn
\<m|\partial_t n\> &=& {\<m|\partial_t \hat H_0(\lambda(t)) |n\> \over E_n - E_m},
\label{c1compact}
\end{eqnarray}
where $|m\>, |n\>$ are short hand notation for $|m(t)\>, |n(t)\>$, and we further assume the system is non-degenerated $( E_n - E_m\neq 0)$. And (\ref{c1coefficient}) becomes
\begin{equation}
\dot C_m = -C_m \<m|\partial_t m\> -\sum\limits_{n\neq m}C_n {\<m|\partial_t \hat H_0(\lambda(t)) |n\> \over E_n - E_m}e^{{i}\theta_n}.
\end{equation}
When  $ \dot \lambda\rightarrow 0$(compared with the level spacing $E_n - E_m$),
\begin{equation}
{\partial_t \hat H_0(\lambda(t)) \over E_n - E_m} = {\dot \lambda \ \partial_\lambda \hat H_0(\lambda) \over E_n - E_m} \approx 0,
\end{equation}
thus
\begin{equation}
\dot C_m \approx -C_m \<m|\partial_t m\> .
\end{equation}
Notice $\<m|\partial_t m\>$ is purely imaginary, as $ \<m|\partial_t m\> +\<\partial_t m| m\> = 0$, we immediately know
${d \over dt} |C_m|=0$, i.e. {\it adiabatic approximation} holds. This approximation means there is no transition between eigenstates of $\hat H_0(\lambda(t)) $. And slowly changing parameter is the common approach.

\subsection{Adiabatic Condition for Quantum Ensemble}
In the previous derivations, we have shown that if $\lambda$ changes slowly, adiabatic approximation holds, i.e. the probability of a state falling in a certain instantaneous eigenstate of $\hat H_0(\lambda(t))$ is constant. This implies that there is no transition between states since probability is conserved. However, if the initial state to be a mixed state $\rho$,
\begin{equation}
\rho(0) = \sum\limits_{n} P_n |n(0)\>\<n(0)|,
\end{equation}
 where $P_n \ge 0$ and $\sum_{n}P_n =1$. If adiabatic approximation holds, it is easy to conclude that the state at time $t$ will be
\begin{equation}
\rho(t) = \sum\limits_{n} P_n |n(t)\>\<n(t)|,
\end{equation}
and the probability of falling in $ |n(t)\>$ is a still constant. In other words, for an quantum ensemble undergoing quantum adiabatic process, the population on each energy level will not change as time evolves.

Here we would like to distinguish the above quantum adiabatic process from the semistatic adiabatic process in thermodynamics. Consider a Gibbs canonical ensemble with partition function $Z(0) = Tr(e^{-\beta \hat H_0(\lambda(0))})$,
\begin{equation}
\rho(0) \equiv {1\over Z(0)}e^{-\beta \hat H_0(\lambda(0))} =  \sum\limits_{n} {e^{-\beta E_n(0)}\over Z(0)} |n(0)\>\<n(0)|,
\label{c2gibbs}
\end{equation}
thus a quantum adiabatic process turn the state into
\begin{equation}
\rho(t) = \sum\limits_{n} {e^{-\beta E_n(0)}\over Z(0)} |n(t)\>\<n(t)|.
\end{equation}
Meanwhile, for a Gibbs\rq{} ensemble undergoing a seimistatic adiabatic process, the system is always in equilibrium, hence
\begin{equation}
\rho\rq{}(t) = {1\over Z(t)}e^{-\beta \hat H_0(\lambda(t))}= \sum\limits_{n} {e^{-\beta E_n(t)}\over Z(t)} |n(t)\>\<n(t)|,
\end{equation}
where  $Z(t) = Tr(e^{-\beta \hat H_0(\lambda(t))})$ is the partition function when parameter $\lambda$ is $\lambda(t)$. In general, since the energy level spacing between states is not fixed, $\rho(t)\neq \rho\rq{}(t)$, and the quantum adiabatic process discussed in the adiabatic theorem is NOT the equilibrium thermodynamical adiabatic process. The canonical ensemble in classical adiabatic theory encounters the same problem. In the rest of this article, when we use \lq\lq{}adiabatic process\rq\rq{}, we are actually referring to the adiabatic process described by either quantum or classical adiabatic theorem, rather than the equilibrium thermodynamical one.

%%%%%%%%%%%%%%%%%%%%%%%%%%%%%%%%
\section{Classical Adiabatic Theorem}
Classical adiabatic theorem is based on a special set of canonical coordinates called action-angle variables. In this section we will start with canonical transformation, followed by a brief introduction about the mathematics of the action-angle variables. We will also make analogues between action variable and quantum number to clarify the relationship between classical and quantum adiabatic theorems.

\subsection{Action-Angle Variable}
Given a time-dependent Hamiltonian $H_0(p,q,\lambda(t))$, let $E = H_0(p,q,\lambda(t))$ representing the system energy. The the action-angle is defined by
\begin{eqnarray}
I \equiv {1\over 2\pi}\oint pdq = {1\over 2\pi}\oint p(q,\lambda,E)dq.
\label{c1action}
\end{eqnarray}
The integral is integrated over a periodic $q$. For a system has more degrees of freedom, the existence of such integral requires the system to be separable \cite{Gong}, hence action-angle variable might not exist for some systems.
If action-angle variables exist, (\ref{c1action}) implies that $E$ is independent of $\theta$,
\begin{eqnarray}
I ={1\over 2\pi}\oint p(q,\lambda,E)dq\ \ \Rightarrow \ \ E = \tilde H_0(I,\lambda) =  H_0(p,q,\lambda).
\label{c1cyclic}
\end{eqnarray}
Notice that although $H_0$ and $\tilde H_0$ has the same value, they have different dependent variables, thus we use tilde to distinguish one from the other.

The advantage of action-angle variable is, if the Hamiltonian is time-independent ($\lambda(t)=const$), $\tilde H_0(I,\lambda)$ becomes our new Hamiltonian under $(I,\theta)$. Remember $\tilde H_0(I,\lambda)$ is independent of $\theta$, we have
\begin{equation}
\dot I = -{\partial \tilde H_0 \over \partial \theta} = 0,
\label{c1independent}
\end{equation}
i.e. $I$ is constant as time evolves.

Now let us come back to the physical interpretation of $I$. Equation (\ref{c1action}) reminds us of {\it Bohr-Sommerfeld quantization} in old quantum theory, which obeys
\begin{equation}
\oint\limits_{H(p,q)=E} p_i dq_i = n_i h,
\end{equation}
where $n_i$ are quantum numbers in old quantum theory. A famous result is the quantization of angular momentum $L$ in Bohr model,
\begin{eqnarray}
\oint\limits_{H(p,q)=E} p dq\ =\ 2\pi r p &=& 2\pi L\ = \ n h, \nn
L &=& n \hbar.
\end{eqnarray}

One important property of quantum adiabatic theorem is that an energy eigenstate remains on the corresponding instantaneous energy eigenstate of the system, or in other words, the quantum number $n$ is invariant. So it is quite natural to require
\begin{equation}
\dot I \approx 0
\end{equation}
in classical adiabatic theorem, when $H_0(p,q,\lambda(t))$ is changing slowly.

\subsection{Adiabatic Approximation}
In order to derive classical adiabatic theorem, we need to know the new Hamiltonian $K_0(I,\theta,t)$ when the coordinate is changed from $(p,q)$ to $ (I,\theta)$. According to (\ref{c1independent}), when $H_0$ is time independent,  $K_0 = \tilde H_0(I,\lambda = const)$, which could greatly simplify our calculation. However, to carry out the transformation under time-dependent $H_0(p,q,\lambda(t))$, we need to find the so-called type-II generating function $F_2(I,q,\lambda)$ \cite{Gold}. Here we will skip this step, and show the subsequent steps. Explicit example in 1-D harmonic oscillator will be shown in the next chapter.

Once we get the generating function  $F_2(I,q,\lambda)$, the relations between coordinates $(p,q)$ and$ (I,\theta)$ are given by
\begin{eqnarray}
p = {\partial F_2(I,q,\lambda)\over \partial q}, \nn
\theta =  {\partial F_2(I,q,\lambda)\over \partial I }.
\label{c1generate}
\end{eqnarray}
In principle, with the above equations, we can solve $(p,q)$ as functions of $(I,\theta)$, i.e. $p(I,\theta,\lambda)$ and $q(I,\theta,\lambda)$. Hamiltonian $H_0(p,q,\lambda(t))$ in $(p,q)$ coordinate and $K_0(I,\theta,t)$ in $(I,theta)$ coordinate are related by
\begin{eqnarray}
K_0(I,\theta,t) &=& \left[H_0(p,q,\lambda) +\left( {\partial F_2(I,q,\lambda)\over \partial t}\right)\Biggr|_{I,\ q}\right]\Biggr|_{p=p(I,\theta,\lambda),\ q=q(I,\theta,\lambda)} \nn
                     &=& H_0(p(I,\theta,\lambda),q(I,\theta,\lambda)),\lambda) + \left[\left( {\partial F_2(I,q,\lambda)\over \partial t}\right)\Biggr|_{I, q}\dot \lambda\right]\Biggr|_{q=q(I,\theta,\lambda)} \nn
                     &=& \tilde H_0(I,\theta,\lambda) + \left[\left( {\partial F_2(I,q,\lambda)\over \partial t}\right)\Biggr|_{I,\ q}\dot \lambda\right]\Biggr|_{q=q(I,\theta,\lambda)}
\end{eqnarray}
From (\ref{c1cyclic}) we know $\tilde H_0$ is independent of $\theta$, therefore we obtain
\begin{equation}
K_0(I,\theta,t) =\tilde H_0(I,\lambda) + \left[\left( {\partial F_2(I,q,\lambda)\over \partial t}\right)_{I,\ q}\dot \lambda\right]\Biggr|_{q=q(I,\theta,\lambda)}
\label{c1K}
\end{equation}
The dynamics of $I$ is given by
\begin{eqnarray}
 \dot I &=& -{\partial K_0(I,\theta,t)\over \partial \theta} \nn
 &=&0 -  {\partial \over \partial \theta}\left(\left[{\partial F_2(I,q,\lambda)\over \partial \lambda}\dot \lambda\right]\Biggr|_{q=q(I,\theta,\lambda))}\right)\Biggr|_{I\ const}  \nn
 &=& -\dot \lambda {\partial \over \partial \theta} \left(\left[{\partial F_2(I,q,\lambda)\over \partial \lambda}\dot \lambda\right]\Biggr|_{q=q(I,\theta,\lambda))}\right)\Biggr|_{I\ const}
 \label{c1Idot}
\end{eqnarray}
Here, if we cheat a little bit, we could claim that $\dot I \approx 0$ when $\dot\lambda$ approached 0. In addition, we notice that if $H_0$ is time-independent, i.e. $\dot \lambda = 0$, equation (\ref{c1Idot}) is reduced to $\dot I = 0$, which is consistent with (\ref{c1independent}).

In fact, the classical adiabatic theorem requires one more step. This step is used to guarantee when $\dot \lambda \approx 0$, (\ref{c1Idot}) does not result in significant change of $I$ as time accumulates. To prove this, we actually require $\dot\lambda/\lambda \ll \omega$, where $\omega$ is the inherent angular frequency of the system. An complete deviation will be given in Goldstein\rq{}s textbook \cite{Gold}.

%%%%%%%%%%%%%%%%%%%%%%%%%%%%%%%%%
\chapter{Work Function}
%%%%%%%%%%%%%%%%%%%%%%%%%%%%%%%%%%%
Work function is an important concept for small systems. It is defined as the probability distribution of the work done to the system during a certain process.

One may be wondering why there is distribution of work. As we know, in general, for a thermodynamical system, the macroscopic quantities have fluctuations. However, due to the large number of particles ($10^{23}$) in large systems, the fluctuation is negligible, and the probability distribution of the macroscopic quantities is quite close to a $\delta$-function. But the case is quite different for small systems. For example, for a canonical ensemble, the ratio between the fluctuation and average of the system energy is given \cite{Huang} by
\begin{eqnarray}
{\sqrt{\<E^2\>-\<E\>^2 }\over \<E\>} \propto {1\over \sqrt {N}},
\end{eqnarray}
where $N$ is the particles in the system. For a system whose $N$ is large enough, the fluctuation is negligible. In contrary, for small systems, the macroscopic quantity spreads in a wide range. In order to describe the work precisely, we need to take work distribution and work fluctuation into consideration.

This chapter mainly discusses the work function and work fluctuation. We will introduce work function under both classical and quantum scheme. In each section, we will first give the definition of work and the general form of work function. At last we will briefly introduce Jarzynski equality.

%%%%%%%%%%%%%%%%%%%%%%%%%%%%%%%%%%%
\section{Work Function for Classical Ensemble}
Before defining the work function, we should make it clear what is the work done to the system in classical mechanics. In this project we will follow Jarzynski approach of inclusive work \cite{HanReview}\cite{Jar}. Notice we always assume the system is not in contact with heat reservoir during the whole process,so the work is simply the energy difference of final and initial state.

Assume time $t$ varies from $0$ to $\tau$, and Hamiltonian $H(p,q,t)$ is given, then in principle we could solve for the trajectory of the system,
\begin{eqnarray}
p(t) = p(p_0,q_0,t), \ \ \ \ q(t) = q(p_0,q_0,t).
\end{eqnarray}
Here the system is not restricted to 2 degrees of freedom, and $p,\ q$ can have arbitrary many components. Given any initial condition $(p_0,q_0)$ of the system, during the action time $\tau$ of the process, work $W$ can be worked out as a function $W_\tau$ of $(p_0,q_0)$,
\begin{eqnarray}
W_\tau(p_0,q_0) = H(p(p_0,q_0,\tau),q(p_0,q_0,\tau),\tau) - H(p_0,q_0,0).
\label{c1solution}
\end{eqnarray}
This equation is totally deterministic. For the system to have fluctuation on work $W$, the initial condition must be given in the form of classical ensemble $\rho(p_0,q_0)$.  $\rho(p_0,q_0)$ is the probability distribution of initial condition in phase space. Then the probability density $P(W)$ of work satisfies
\begin{equation}
P(W)dW = \int\limits_{W\leq W_\tau(p_0,q_0)<W+dW}\rho(p_0,q_0)dp_0dq_0,
\end{equation}
or equivalently,
\begin{equation}
P(W) =\int\limits_{\Gamma}\rho(p_0,q_0)\delta(W-W_\tau(p_0,q_0))dp_0dq_0.
\label{c3classwork}
\end{equation}
$\delta(x)$ is the Dirac-delta function, and $\Gamma$ under the integral means that the integration is performed over the entire phase space $\Gamma$.

%%%%%%%%%%%%%%%%%%%%%%%%%%%%%%%
\section{Work Function for Quantum Ensemble}
For quantum ensemble, the case is slightly different. Let $E_n(0)$ be the eigenenergies of the system at $t = 0$, similarly for $E_n(\tau)$. First consider a system starts with a pure state $|n(0)\>$, there are several final states the system might fall in. The transition probability of initial state $|n(0)\>$ to final state $|m(\tau)\>$ is
\begin{equation}
P_{n\rightarrow m} \equiv  |\<m(\tau)|\hat U(\tau,0)|n(0)\>|^2,
\end{equation}
where $\hat U(\tau,0)$ is the time-evolution from 0 to $\tau$. The work done is simply $E_m(\tau)-E_n(0)$. Thus for this single state $|n(0)\>$, the work function is discretized,
\begin{equation}
P_n(W) = \sum\limits_{m}P_{n\rightarrow m}\delta \Bigl( W-[E_m(\tau)-E_n(0)]\Bigr).
\end{equation}
It is easy to verify that $P_n(W)$ is normalized.
Similarly, let\rq{}s now consider a quantum ensemble, i.e. a mixed state
\begin{equation}
\rho(0) = \sum\limits_{n}P_n|n(0)\>\<n(0)|
\end{equation}
In order to measure the work done to the system, we need to measure the system energy at both $t = 0$ and $t = \tau$ to get the exact system energy difference. In mathematics, this is equivalent to projecting the mixed state onto instantaneous energy eigenstates of $\hat H_0(t)$. The corresponding $P(W)$ is
\begin{equation}
P(W) = \sum\limits_{n}\sum\limits_{m}P_n P_{n\rightarrow m}\delta \Bigl( W-[E_m(\tau)-E_n(0)]\Bigr).
\label{c1quantumwf}
\end{equation}

%%%%%%%%%%%%%%%%%%%%%%%%
\section{Jarzynski Equality}
Jarzynski equality is an important equation relating non-equilibrium and equilibrium process under a fixed temperature. A brief idea of the classical theory is stated below. Remember we define work in Jarzynski\rq{}s approach (\ref{c1solution}), i.e.,
\begin{eqnarray}
W_\tau(p_0,q_0) = H(p(p_0,q_0,\tau),q(p_0,q_0,\tau),\tau) - H(p_0,q_0,0),
\end{eqnarray}
for a certain trajectory with $(p_0,q_0)$ as initial condition. Thus for a system start with a Gibbs canonical ensemble
\begin{equation}
\rho(p_0,q_0,0) = {e^{-\beta H(p_0,q_0,0)}\over Z_0},
\end{equation}
the expectation
\begin{eqnarray}
\<e^{-\beta W}\> &=& \int\limits_{\Gamma}{e^{-\beta H(p_0,q_0,0)}\over Z_0}e^{-\beta W_\tau(p_0,q_0)}dp_0dq_0 \nn
                               &=&\int\limits_{\Gamma}{e^{-\beta H(p_0,q_0,0)}\over Z_0}e^{-\beta (H(p(p_0,q_0,\tau),q(p_0,q_0,\tau),\tau) - H(p_0,q_0,0))}dp_0dq_0   \nn
                               &=&{1\over Z_0}\int\limits_{\Gamma}e^{-\beta H(p(p_0,q_0,\tau),q(p_0,q_0,\tau),\tau) }dp_0dq_0   \nonumber.
\end{eqnarray}
Notice that $(p_0,q_0) \rightarrow (p(p_0,q_0,\tau),q(p_0,q_0,\tau))$ is a canonical transformation, thus the Jacobian is equal to $1$, and
\begin{eqnarray}
  \<e^{-\beta W}\>    &=&{1\over Z_0}\int\limits_{\Gamma}e^{-\beta  H(p(p_0,q_0,\tau),q(p_0,q_0,\tau),\tau) }dp(p_0,q_0,\tau)dq(p_0,q_0,\tau) \nn
                                    &=&{Z_\tau\over Z_0},
                                    \label{c3interJar}
\end{eqnarray}
where $Z_\tau$ is the partition function when $\lambda=\lambda(\tau)$. Furthermore, we know the relation between Helmholtz free energy and partition function,
\begin{equation}
F = -{1\over \beta}\ln Z, \ \ \ \ {\rm or}\ \ \ \ Z = e^{-\beta F}
\end{equation}
plug into (\ref{c3interJar}), and Jarzynski equality emerges,
\begin{equation}
\<e^{-\beta W}\> = {e^{-\beta F_\tau}\over e^{-\beta F_0}} = e^{-\beta \Delta F}.
\label{c3Jar}
\end{equation}
This equation is very powerful in the sense that it relates non-equilibrium quality $W$ with equilibrium quality $F$, with regardless of work function $P(W)$. In practical aspect, if we want to measure the free energy difference between two equilibrium state, we only need to prepare the Gibbs canonical ensemble, randomly pick a sample from it, and change the parameter $\lambda$ from $\lambda(0)$ to $\lambda(\tau)$ (might be very fast). We do not even need to wait for the final state turning to equilibrium. As if we repeat such non-equilibrium process for enough times and measure the work done during the process, we could estimate $\Delta F$ through Jarzynski equality. For example, if we hope to measure the free energy increase when the length of a protein is changed $\Delta l$, a conventional way is stretching the protein very slowly, so that the process can be regarded as semi-static. Such process costs much time. But with Jarzynski equality, we need only to stretch the protein by length $\Delta l$ and measure the work $W$ in this process. Average of $e^{-\beta W}$ will give us estimation of $\Delta F$.

However, we notice that (\ref{c3Jar}) only contains information about the average, and  it does not tell us information on work fluctuation. For some system, the work fluctuation might be very large, and the expectation value $\<e^{-\beta W}\>$ converges slowly. Later we will show how fast-forward adiabatic process helps $\<e^{-\beta W}\>$ to converge faster, by shrinking the work fluctuation.

%%%%%%%%%%%%%%%%%%
\chapter{Classical Fast-Forward Adiabatic Process}
%%%%%%%%%%%%%%%%%%
In the introduction chapters we have briefly discussed the pros and cons of the conventional adiabatic process. Conventional adiabatic process suppresses the work fluctuation significantly according to Lutz \cite{Lutz}. On the other hand, it takes comparatively longer time since parameter must change slowly enough. So far there is no literal discussion focused on overcoming such difficulty in classical cases. However, our pioneering work on fast-forward adiabatic process fills in this gap. It does not only boost up the speed, but also suppresses the work fluctuation compare with non-adiabatic process with the same speed. This chapter will explain the details of our original work in classical theory.

This chapter is aimed to show how to guarantee action $I$ invariant even if the parameter changes fast. Basically we will ensure this by adding a control field $H_C$ onto the original Hamiltonian $H_0$.  An explicit example of 1-D classical harmonic oscillator will be shown for different processes. We will then compare their work functions, hence show how the control field $H_C$ suppresses the work fluctuation of a Gibbs canonical ensemble. Simulation results will also be shown at the end of the chapter.
\section{Control Field $H_C$}
\subsection{Formal Solution}
Assume we have obtained the type-II generating function $F_2(I,q,\omega)$, we can immediately get the relation between $(p,q)$ and $(I,\theta)$. According to (\ref{c1K}) and (\ref{c1Idot})
\begin{equation}
K_0(I,\theta,t) =\tilde H_0(I,\lambda) + \left[\left( {\partial F_2(I,q,\lambda)\over \partial t}\right)\Biggr|_{I,\ q}\dot \lambda\right]\Biggr|_{q=q(I,\theta,\lambda)}
\label{c2K}
\end{equation}
and
\begin{equation}
 \dot I ={\partial K_0(I,\theta,t)\over \partial \theta} =  -\dot \lambda {\partial \over \partial \theta} \left(\left[{\partial F_2(I,q,\lambda)\over \partial \lambda}\dot \lambda\right]\Biggr|_{q=q(I,\theta,\lambda))}\right)\Biggr|_I\approx 0,
 \label{c2Idot}
\end{equation}
when $\dot \lambda\ll 1$ adiabatic approximation holds. However, if $\dot \lambda$ is not negligible, (\ref{c2Idot}) is the only term term which might change the value of $I$. This is resulted from the $\theta$ dependence of the second term in (\ref{c2K}).

Our method of speeding up the adiabatic process (keep $I$ constant) is very straight forward. We will add a control field $K_C$ to $K_0$, such that the new Hamiltonian $K = K_0 +K_C$ is $\theta$ independent, which leads to
\begin{equation}
 \dot I ={\partial K(I,\theta,t)\over \partial \theta} =0.
\end{equation}
One obvious solution is
\begin{equation}
K_C(I,\theta,t) =- \left[\left( {\partial F_2(I,q,\lambda)\over \partial t}\right)\Biggr|_{I,\ q}\dot \lambda\right]\Biggr|_{q=q(I,\theta,\lambda)},
\end{equation}
so that
\begin{equation}
K(I,\theta,t) = K_0(I,\theta,t) +K_C(I,\theta,t) = \tilde H_0(I,\lambda),
\end{equation}
which is $\theta$ independent. In explicit examples, if necessary, we could transform it back to $(p,q)$ coordinate to get a more familiar physical picture.
\subsection{Gibbs Canonical Ensemble}
Since we are much concerned about the work function, it is necessary for us to specify which ensemble we are working with. We will use Gibbs canonical ensemble as our start point, because it is the most natural ensemble to deal with and also simple to prepare in experiment. Then for this classical ensemble, the partition function at $t = 0$ is
\begin{equation}
Z_0 = \int\limits_{\Gamma} e^{-\beta H(p_0,q_0,\lambda(0))}dp_0dq_0,
\end{equation}
where $\beta = {1/k_B T}$ is the conventional inverse temperature. Distribution of initial momentum $\p_0$ and position $q_0$ is
\begin{equation}
\rho(p_0,q_0,0) = {e^{-\beta H(p_0,q_0,\lambda(0))}\over Z_0}.
\end{equation}
By (\ref{c3classwork}), the work function is
\begin{equation}
P(W) =\int\limits_{\Gamma}{e^{-\beta H(p_0,q_0,\lambda(0))}\over Z_0}\delta(W-W_\tau(p_0,q_0))dp_0dq_0.
\label{c2classicalwf}
\end{equation}

%%%%%%%%%%%%%%%%%%%%%%%%%%%%%
\section{Application in 1-D Classical Harmonic Oscillator}
In this section we will show the results for 1-D classical harmonic oscillator. The original Hamiltonian $H_0$ is given by
\begin{equation}
H_0(p,q,\omega(t)) = {p^2\over 2m} +{1\over 2}m\omega^2(t)q^2.
\end{equation}
where $\omega(t)$ plays the role of $\lambda(t)$. We will first derive $H_C$ and work function under a certain $\omega(t)$. We will then give the work function under a process without $H_C$. Adiabatic and sudden limit will also be given for comparison.
\subsection{Fast-Forward Adiabatic Process}
First calculate action $I$. Follow (\ref{c1action}) and (\ref{c1cyclic}), let $E = H_0(p,q,\omega(t))$,
\begin{eqnarray}
I &\equiv& {1\over 2\pi}\oint pdq ={1\over \pi} \int\limits_{q_{min}}^{q_{max}}\sqrt{2mE - m^2\omega^2q^2}dq \nn
   &=&{1\over \pi}\sqrt{2mE}\int\limits_{q_{min}}^{q_{max}}\sqrt{1 - {m^2\omega^2\over2mE}q^2}dq \nn
   &=& {2mE\over \pi m\omega}\int\limits_{-1}^{1}\sqrt{1 - s^2}ds \nn
    &=& {E\over \omega}
\end{eqnarray}
Or simply $\tilde H_0(I,\omega)=E = \omega I$

As illustrated previously, we also need to find the type-II generating function $F_2(I,q,\omega)$. As (\ref{c1generate}) indicates,
\begin{eqnarray}
p = {\partial F_2(I,q,\omega)\over \partial q},
\end{eqnarray}
thus
\begin{eqnarray}
\omega I =E &=& {1\over 2m}\left({\partial F_2(I,q,\omega)\over \partial q}\right)^2+{1\over  2}m\omega^2 q^2 \nn
{\partial F_2\over \partial q} &=& \sqrt{2m\omega I - m^2\omega^2 q^2} \nn
F_2(I,q,\omega) &=& \int \sqrt{2m\omega I - m^2\omega^2 q^2}dq
\end{eqnarray}
and $\theta$ is given by
\begin{eqnarray}
\theta &=&  {\partial F_2(I,q,\omega)\over \partial I} \nn
       &=&\int {\partial \over \partial I}\sqrt{2m\omega I - m^2\omega^2 q^2}dq \nn
       &=&\int {m\omega \over \sqrt{2m\omega I - m^2\omega^2 q^2}}dq \nn
       &=&\int \sqrt{m\omega\over 2I}{dq\over \sqrt{1 - {m\omega\over 2I} q^2}} \nn
       &=&\arcsin(\sqrt{m\omega\over 2I} q),
\end{eqnarray}
or
\begin{equation}
q =  \sqrt{2I\over m\omega}\sin\theta.
\end{equation}
By (\ref{c1K})
\begin{eqnarray}
K_0(I,\theta,t) &=&\tilde H_0(I,\lambda) + \left[\left( {\partial F_2(I,q,\omega)\over \partial t}\right)_{I,\ q}\dot \omega\right]\Biggr|_{q=q(I,\theta,\omega)} \nn
         &=&\omega I + \dot \omega\left[ \int {\partial \over \partial \omega}\sqrt{2m\omega I - m^2\omega^2 q^2}dq\right] \Biggr|_{q=q(I,\theta,\omega)} \nn
         &=&\omega I + \dot \omega\left[\int {(m I -m^2 \omega q^2)dq\over \sqrt{2m\omega I - m^2\omega^2 q^2}}\right] \Biggr|_{q=q(I,\theta,\omega)} \nn
         &=&\omega I + \dot \omega\left[\int {mI\over \sqrt{2m\omega I}}{1-{m\omega\over I} q^2 dq\over \sqrt{1 - {m\omega\over 2I} q^2}}\right] \Biggr|_{q=q(I,\theta,\omega)}  \nn
         &=&\omega I + \dot \omega\left[\int {I\over\omega}{1-{m\omega\over I} q^2 d\sqrt{m\omega\over 2I}q\over \sqrt{1 - {m\omega\over 2I} q^2}}\right] \Biggr|_{q=q(I,\theta,\omega)}  \nonumber
\end{eqnarray}
substitute $ \sqrt{m\omega\over 2I}q $ with $\sin s$,
\begin{eqnarray}
K_0(I,\theta,t) &=&\omega I+{\dot \omega I\over \omega}\left[\int{1-2\sin^2s d\sin s\over \cos s}\right] \Biggr|_{q=q(I,\theta,\omega)} \nn
                     &=&\omega I+{\dot \omega I\over \omega}\left[\int{(1-2\sin^2s) d s}\right] \Biggr|_{q=q(I,\theta,\omega)} \nn
                     &=&\omega I+{\dot \omega I\over \omega}\left[\int{\cos(2s) d s}\right] \Biggr|_{q=q(I,\theta,\omega)} \nn
                     &=&\omega I+{\dot \omega I\over 2\omega}\left[\sin (2\arcsin(\sqrt{m\omega\over 2I}q) )\right] \Biggr|_{q=q(I,\theta,\omega)} \nn
 ({\rm plug\ in \ }q =  \sqrt{2I\over m\omega}\sin\theta)\ \ \  &=&\omega I+{\dot \omega I\over 2\omega}\sin(2\theta).
\label{c2Kfinal}
\end{eqnarray}
Thus an obvious control field $K_C$ to make $K = K_0 + K_C$ $\theta$ independent is
\begin{equation}
K_C(I,\theta,t) = -{\dot \omega I\over 2\omega}\sin(2\theta) = -{\dot \omega\over \omega}I\sin\theta \cos\theta.
\end{equation}
And as we know,
\begin{equation}
p = \sqrt{2mE - m^2\omega^2 q^2} =\sqrt{2m\omega I -  m^2\omega^2 {(\sqrt{2I\over m\omega}\sin\theta)}^2} =\sqrt{2m\omega I}\cos\theta
\end{equation}
so control field in $(p,q)$ coordinate is
\begin{equation}
H_C(p,q,t)=K_C(I,\theta,t) = - {\dot \omega\over 2\omega} p q.
\end{equation}
Before working out the work function $P(W)$ explicitly, in order to compare the fast-forward adiabatic process with other processes, we hope to choose our $\omega(t)$ such that $H = H_0 +H_C$ reduces to $H_0$ at the start ($t=0$) and the end ($t=\tau$) of the process, i.e. $H_C(p,q,0) =H_C(p,q,\tau)=0 $. We will always assume this requirement is satisfied.
With such $\omega$, the canonical ensemble $\rho$ (or, our initial probability distribution of $(p_0,q_0)$) is given by
\begin{equation}
\rho(p_0,q_0,0) = {e^{-\beta H(p_0,q_0,\omega(0))}\over Z},
\end{equation}
which implies
\begin{equation}
\rho(I_0,\theta_0,0) = {e^{-\beta \omega_0 I_0}\over Z_0}\biggr|{\partial(I,\theta) \over \partial (p,q)}\biggr|={e^{-\beta \omega_0 I_0}\over Z_0}.
\label{c2workfunction1}
\end{equation}
Here ${\partial(I,\theta) \over \partial (p,q)}$ is the Jacobian matrix, and its determinate equals to $1$ since the transformation is canonical. And partition function at $t=0$ is
\begin{eqnarray}
Z_0 &=&\int\limits_{\Gamma} e^{-\beta H(p_0,q_0,\omega(0))}dp_0dq_0         \nn
       &=&\int\limits_{\Gamma} e^{-\beta \omega_0 I_0}dI_0d\theta_0 = {2\pi\over \beta \omega_0}.
       \label{c2workfunction2}
\end{eqnarray}
Since $H_C = 0$ at $t = 0,\tau$, the work done to the system is
\begin{eqnarray}
W_\tau(I_0,\theta_0) &=&H(I(I_0,\theta_0,\tau),\theta(I_0,\theta_0,\tau),\tau)-H(I_0,\theta_0,0) \nn
                                &=& H_0(I(I_0,\theta_0,\tau),\theta(I_0,\theta_0,\tau),\tau)-H_0(I_0,\theta_0,0) \nn
       (I = const, I(\tau)=I(0)=I_0)                        &=&\omega(\tau) I_0 - \omega(0) I_0 \equiv \Delta\omega I_0.
       \label{c2workfunction3}
\end{eqnarray}
Although $H_C$ vanished at both end, it still does work to the system. So here the work is the total work of both $H_0$ and $H_C$. Next, plug (\ref{c2workfunction1}), (\ref{c2workfunction2}) and (\ref{c2workfunction3}) into (\ref{c2classicalwf}),
\begin{eqnarray}
P(W) &=&\int\limits_{\Gamma}{e^{-\beta H(p_0,q_0,\omega(0))}\over Z_0}\delta(W-W_\tau(p_0,q_0))dp_0dq_0 \nn
     &=&\int\limits_{\Gamma}{\beta \omega_0\over 2\pi}{e^{-\beta \omega_0 I_0}}\delta(W-\Delta\omega I_0)dI_0d\theta_0 \nn
     &=&{\omega_0\beta\over\Delta\omega}\exp(-{\omega_0\over\Delta\omega}\beta W),
\label{c21Drho}
\end{eqnarray}
which is an exponential distribution. Both the expectation and standard deviation are ${\Delta\omega\over\omega_0\beta}$.

\subsection{Finite-Time Process}
As we know, conventional adiabatic process has infinitely long duration $\tau$ as the parameter is changing infinitely slow. We now consider finite-time process with arbitrary finite duration $\tau$. By adiabatic theorem, such finite-time process can be reduced to conventional adiabatic process as if we choose a sufficiently long duration $\tau$.

First of all, although $H_C$ is not used in this section, to compare work with fast forward adiabatic process, we hope $\omega(t)$ could lead to $H_C(0) = H_C(\tau) =0$. Such that
\begin{equation}
W_\tau =H(\tau)-H(0) = H_0(\tau)+H_C(\tau)-H_0(0)-H_C(0)=H_0(\tau)-H_0(0),
\end{equation}
i.e. the definition work is consistent. One way of ensuring this is
\begin{equation}
\omega(t) =\omega_0 \sqrt{{f^2+1\over 2}-{f^2-1\over2}\cos (n\pi{t\over\tau})},
\label{c4omega}
\end{equation}
where $f$ is a real number and $n$ is an integer. It is easy to verify that $\dot\omega(0)=\dot\omega(\tau)=0$, which implies $H_C(p,q,0) =H_C(p,q,\tau)=0 $. In the context, if not specify we will choose $n=1$ for convenience. The advantage of choosing $n=1$ is that
\begin{equation}
\omega(0) =\omega_0, \ \ \ \ \omega_{f} \equiv\omega(\tau) = f \omega_0.
\end{equation}
Actually $f$ stands for factor of $\omega$ being increased.

Under such realization of $\omega(t)$, the dynamics is
\begin{eqnarray}
\dot p = -{\partial H_0 \over \partial q} &=& -m\omega^2(t) q = -m\omega_0^2\left[{f^2+1\over 2}-{f^2-1\over2}\cos (\pi{t\over\tau})\right]q,\label{c2pdot} \\
\dot q &=& {\partial H_0 \over \partial p} = {p\over m} .
\label{c2qdot}
\end{eqnarray}
Differentiate (\ref{c2qdot}) with respect to $t$ and plug it into (\ref{c2pdot})
\begin{equation}
\ddot q(t) + \omega_0^2\left[{f^2+1\over 2}-{f^2-1\over2}\cos (\pi{t\over\tau})\right]q(t) =0.
\label{c2diffeqn1}
\end{equation}
Let $\pi t/\tau = 2x$, or  $ t = 2\tau x/\pi$ we get
\begin{equation}
{\pi^2\over 4\tau^2}{d^2\over dx^2} q(x) + \omega_0^2\left[{f^2+1\over 2}-{f^2-1\over2}\cos (2x)\right]q(x) =0,
\end{equation}
or
\begin{equation}
{d^2\over dx^2} q(x) + {4\tau^2\omega_0^2\over \pi^2}\left[{f^2+1\over 2}-{f^2-1\over2}\cos (2x)\right]q(x) =0.
\end{equation}
Then define
\begin{equation}
a \equiv {4\tau^2\omega_0^2\over \pi^2}{f^2+1\over 2}\ \ {\rm and}\  \ b  \equiv {1\over 2}{4\tau^2\omega_0^2\over \pi^2}{f^2-1\over2},
\end{equation}
we get
\begin{equation}
{d^2\over dx^2} q(x) + [a-2b\cos(2x)]q(x) =0,
\label{c2diffeqn2}
\end{equation}
which is exactly {\it Mathieu\rq{}s differential equation}. The independent solutions are called {\it Mathieu sin} and
{\it Mathieu cos}, which satisfies
\begin{eqnarray}
\MC(a,b,0) = 1, \ \ &\ & \ \MC\rq{}(a,b,0)=0\ ; \nn
\MS(a,b,0) = 0, \ \ &\ &\ \MS\rq{}(a,b,0)=1.
\label{c2Mathieu}
\end{eqnarray}
Here $\rq{}$ denotes the derivative with respect to $x$. The solution with initial condition $(p_0,q_0)$ is
\begin{eqnarray}
q(t) &=& q_0 \MC(a,b,{\pi\over2\tau}t) + {p_0\over m} {2\tau \over \pi}\MS(a,b,{\pi\over2\tau}t), \nn
p(t) &=& m \dot q(t)=p_0 \MS\rq{}(a,b,{\pi\over2\tau}t) + {\pi\over2\tau} m q_0 \MC\rq{}(a,b,{\pi\over2\tau}t)
\label{c2pq}
\end{eqnarray}
Now let\rq{}s consider the work done to the system. We would like to emphasize that the following arguments hold in general, here we simply use Mathieu function as an example.

Let $C(t)$ and $S(t)$ be solutions of 1-D harmonic oscillator with time-dependent $\omega(t)$. $C$ and $S$ satisfies
\begin{eqnarray}
C(0) = 1, \ \ &\ & \ \dot C(0)=0\ ; \nn
S(0) = 0, \ \ &\ &\ \dot S(0)=1.
\label{S&C}
\end{eqnarray}
Since in our example we have chosen our $\omega(t)$, so the corresponding solutions are
\begin{equation}
C(t) = \MC(a,b,{\pi\over2\tau}t), \ \ \ S(t) = {2\tau \over \pi}\MS(a,b,{\pi\over2\tau}t).
\end{equation}
And solutions with initial condition $(p_0,q_0)$ are
\begin{eqnarray}
q(t) &=& q_0 C(t) + {p_0\over m} S(t), \nn
p(t) &=& m \dot q(t)=p_0 \dot S(t) + m q_0 \dot C(t),
\end{eqnarray}
like (\ref{c2pq}) indicates. Consider (\ref{c1solution}), since we are not considering $H_C$, $H_0(p,q,\omega)$ is our full Hamiltonian,
\begin{eqnarray}
W_\tau(p_0,q_0) &=& H_0(p(p_0,q_0,\tau),q(p_0,q_0,\tau),\tau) - H_0(p_0,q_0,0) \nn
                         &=& {1\over 2m}\left[p_0 \dot S(\tau) + m q_0 \dot C(\tau)\right]^2+{m \omega_{f}^2\over 2}\left[q_0 C(\tau) + {p_0\over m} S(\tau)\right]^2 \nn
                         &\ & -{1\over 2m}p_0^2 - {m \omega_0^2\over 2}q_0^2 \nn
                         &=&K {\beta\over 2m}p_0^2 + L {\beta m\omega_0^2\over 2}q_0^2 +M\beta\omega_0 p_0 q_0,
\label{c2workdone}
\end{eqnarray}
where $\beta$ is the conventional inverse temperature, and
\begin{eqnarray}
K &\equiv&{1\over \beta} {\left[\dot S^2(\tau) +   \omega_{f}^2 S^2(\tau)-1  \right]}, \nn
L &\equiv&{1\over \beta} {\left[{\dot C^2(\tau)\over \omega_0^2} +   {\omega_{final}^2 \over \omega_0^2}C^2(\tau)-1  \right]}, \nn
M &\equiv&{1\over \beta \omega_0} {\left[\dot C(\tau) \dot S(\tau)+   \omega_{f}^2 C(\tau)S(\tau)  \right]}.
\end{eqnarray}
Or in canonical quadratic form,
\begin{equation}
W_\tau(p_0,q_0) =\left(
\begin{array}{cc}
\sqrt{\beta\over 2m}p_0 & \sqrt{\beta m\omega_0^2\over 2}q_0
\end{array}
\right) \left(
\begin{array}{cc}
K & M\\
M & L
\end{array}
\right)
\left(
\begin{array}{c}
\sqrt{\beta\over 2m}p_0\\
\sqrt{\beta m\omega_0^2\over 2}q_0
\end{array}
\right),
\end{equation}
where $K$, $L$, $M$ consist of $S(\tau)$, $\dot S(\tau)$, $C(\tau)$, $\dot C(\tau)$ and system constants, therefore $K$, $L$, $M$ are independent of $p_0,\ q_0$. In our case they contain Mathieu functions, in a more general case we only need to replace the Mathieu function with other special functions, and could always express $W_\tau$ as a canonical quadratic form of $(p_0,q_0)$.

Because the matrix consists of $K$, $L$ and $M$ is symmetric, there exists an orthonormal matrix $O$, s.t.
\begin{equation}
\left(
\begin{array}{cc}
K & M\\
M & L
\end{array}
\right) = O^T \left(
\begin{array}{cc}
\mu_+ &0 \\
0  & \mu_-
\end{array}
\right)O.
\end{equation}
If we define
\begin{equation}
\left(
\begin{array}{c}
p\rq{}\\
q\rq{}
\end{array}
\right) =O\left(
\begin{array}{c}
\sqrt{\beta\over 2m}p_0\\
\sqrt{\beta m\omega_0^2\over 2}q_0
\end{array}
\right) ,
\end{equation}
$W_\tau$ can be expressed as
\begin{eqnarray}
W_\tau &=&\left(
\begin{array}{cc}
p\rq{} & q\rq{}
\end{array}
\right) \left(
\begin{array}{cc}
\mu_+ & 0\\
0 & \mu_-
\end{array}
\right)\left(
\begin{array}{c}
p\rq{} \\ q\rq{}
\end{array}
\right) \nn
&=&\mu_+p\rq{}^2+\mu_-q\rq{}^2
\end{eqnarray}
Doing such complicated transformation helps working out our work function. Notice that $O$ is orthonormal, thus
\begin{eqnarray}
\beta H_0(p_0,q_0,0)&=&\left(
\begin{array}{cc}
\sqrt{\beta\over 2m}p_0&
\sqrt{\beta m\omega_0^2\over 2}q_0
\end{array}
\right) \left(
\begin{array}{c}
\sqrt{\beta\over 2m}p_0\\
\sqrt{\beta m\omega_0^2\over 2}q_0
\end{array}
\right)\nn
&=&  \left(
\begin{array}{cc}
\sqrt{\beta\over 2m}p_0&
\sqrt{\beta m\omega_0^2\over 2}q_0
\end{array}
\right)O^TO \left(
\begin{array}{c}
\sqrt{\beta\over 2m}p_0\\
\sqrt{\beta m\omega_0^2\over 2}q_0
\end{array}
\right) \nn
&=&\left(
\begin{array}{cc}
p\rq{} &
q\rq{}
\end{array}
\right)
\left(
\begin{array}{c}
p\rq{}\\
q\rq{}
\end{array}
\right) = p\rq{}^2+q\rq{}^2
\end{eqnarray}
And work function is given by
\begin{eqnarray}
P(W) &=&\int\limits_{\Gamma}{e^{-\beta H(p_0,q_0,\omega(0))}\over Z_0}\delta(W-W_\tau(p_0,q_0))dp_0dq_0 \nn
     &=&\int\limits_{\Gamma}{e^{-(p\rq{}^2+q\rq{}^2)}\over Z_0}\delta(W-\mu_+p\rq{}^2-\mu_-q\rq{}^2){2\over \beta\omega_0}dp\rq{}dq\rq{} \nn
     &=&\int\limits_{\Gamma}{1\over\pi}e^{-(p\rq{}^2+q\rq{}^2)}\delta(W-\mu_+p\rq{}^2-\mu_-q\rq{}^2)dp\rq{}dq\rq{}.
\end{eqnarray}
Here we use the result $Z_0 = 2\pi /\beta\omega_0$, and additional $2/\beta \omega_0$ factor is the Jacobian from $(p_0,q_0)$ to ${(p\rq{},q\rq{})}$.

To finish the following calculation, we will further assume that matrix consists of $K$, $L$ and $M$ is positive-definite when $\omega_f > \omega_0$, i.e. its two eigenvalues $\mu_+ >\mu_->0$. Because of this, we can alway make the following transformation
\begin{equation}
p\rq{} ={1\over \sqrt{\mu_+}} r \cos \phi, \ \ \ q\rq{} ={1\over \sqrt{\mu_-}} r \sin \phi,
\end{equation}
with $r>0$ and $\phi \in [0,2\pi)$. Easy to calculate the Jacobian from $(p\rq{},q\rq{})$ to $(r,\phi)$ is $r/\sqrt{\mu_+\mu_-}$, thus
\begin{eqnarray}
P(W)  &=&\int\limits_{\Gamma}{1\over\pi}e^{-(p\rq{}^2+q\rq{}^2)}\delta(W-\mu_+p\rq{}^2-\mu_-q\rq{}^2)dp\rq{}dq\rq{}  \nn
     &=&\int\limits_{\Gamma}{1\over\pi}e^{-r^2({\cos^2\phi\over\mu_+}+{\sin^2\phi\over\mu_-})}\delta(W-r^2){r\over\sqrt{\mu_+\mu_-}}drd\phi \nn
     &=&\int_0^{2\pi}{1\over2\pi\sqrt{\mu_+\mu_-}}e^{-W({\cos^2\phi\over\mu_+}+{\sin^2\phi\over\mu_-})}d\phi \nn
     &=&\int_0^{2\pi}{1\over2\pi\sqrt{\mu_+\mu_-}}\exp{\left[-{\mu_++\mu_-\over 2\mu_+\mu_-}W+{\mu_+-\mu_-\over 2\mu_+\mu_-}W\cos(2\phi)\right]}d\phi \nn
       &=&\exp{\left[-{\mu_++\mu_-\over 2\mu_+\mu_-}W\right]}\int_0^{4\pi}{1\over4\pi\sqrt{\mu_+\mu_-}}\exp{\left[{\mu_+-\mu_-\over 2\mu_+\mu_-}W\cos(2\phi)\right]}d2\phi \nn
              &=&4\exp{\left[-{\mu_++\mu_-\over 2\mu_+\mu_-}W\right]}\int_0^{\pi}{1\over4\pi\sqrt{\mu_+\mu_-}}\exp{\left[{\mu_+-\mu_-\over 2\mu_+\mu_-}W\cos(\phi\rq{})\right]}d\phi\rq{} \nn
              &=&{1\over\sqrt{\mu_+\mu_-}}\exp{\left[-{\mu_++\mu_-\over 2\mu_+\mu_-}W\right]}I_0\left[{\mu_+-\mu_-\over 2\mu_+\mu_-}W\right]  \ \ \ \ \ (W\ge 0),
              \label{c4nonadia}
\end{eqnarray}
where $I_0(x)$ is the modified Bessel function of the first kind, with parameter $\alpha = 0$. And we use the formula
\begin{equation}
I_\alpha(x) = \frac{1}{\pi}\int_0^\pi \exp(x\cos\phi\rq{}) \cos(\alpha\phi\rq{}) d\phi\rq{} - \frac{\sin(\alpha\pi)}{\pi}\int_0^\infty \exp(-x\cosh s - \alpha s) ds.
\end{equation}
This is the case for $\omega_f>\omega_0$, and $W\ge 0$. If $\omega_f<\omega_0$, $P(W)$ has a similar expression with $W\le 0$ , except $\mu_\pm$ will be less than $0$.

Although we have got the compact form of work function under finite-time process, it is not easy to calculate $\<W\>$ and higher order moment. And the algebraic relation between $\mu_\pm$ and Mathieu function is quite complicated in our case. Because of this, in later sections we will use numerical and simulation results to compare finite-time process with fast-forward adiabatic process.

\subsection{Adiabatic Process \& Sudden Change}
For conventional adiabatic process, $K(I,\omega) \approx \tilde H_0(I,\omega) =\omega I$. Thus partition function at $t=0$ is
\begin{eqnarray}
Z_0 &=&\int\limits_{\Gamma} e^{-\beta H_0(p_0,q_0,\omega(0))}dp_0dq_0         \nn
       &=&\int\limits_{\Gamma} e^{-\beta \omega_0 I_0}dI_0d\theta_0 = {2\pi\over \beta \omega_0},
\end{eqnarray}
and
\begin{equation}
\rho(I_0,\theta_0,0) = {e^{-\beta \omega_0 I_0}\over Z_0}\biggr|{\partial(I,\theta) \over \partial (p,q)}\biggr|={e^{-\beta \omega_0 I_0}\over Z_0}.
\end{equation}
Since $H_C = 0$ at $t = 0,\tau$, the work done to the system is
\begin{eqnarray}
W_\tau(I_0,\theta_0)&=&\tilde H_0(I(I_0,\theta_0,\tau),\theta(I_0,\theta_0,\tau),\tau)-\tilde H_0(I_0,\theta_0,0) \nn
   &=& \Delta\omega I_0.
\end{eqnarray}
And work function is
\begin{eqnarray}
P(W)    &=&\int\limits_{\Gamma}{\beta \omega_0\over 2\pi}{e^{-\beta \omega_0 I_0}}\delta(W-\Delta\omega I_0)dI_0d\theta_0 \nn
     &=&{\beta\omega_0\over\Delta\omega}\exp(-{\omega_0\over\Delta\omega}\beta W).
     \label{c4convenadw}
\end{eqnarray}
We notice the work function of adiabatic process is identical with (\ref{c21Drho}), the work function of fast-forward adiabatic process. In fact, for any system, if the control field vanishes at the start and the end of the process, the work function should be identical with the conventional adiabatic process. This is due to the $\theta$-independence of $\tilde H_0$, i.e., the initial distribution (Gibbs canonical ensemble $\rho$) and work alone a specific path $W_\tau$ are both independent of $\theta$. In addition, the dynamics of $I$ are the same in both process: $I $ is constant. Thus the work functions are identical.

\

Now let\rq{}s work on system undergoing a sudden change in $\omega$ at $t = 0$. This sudden change condition is an extreme case of non-adiabatic process. Since the duration of this sudden change is infinitely small, the position $q$ and momentum $p$ are not changed, i.e.
\begin{equation}
\lim_{t\rightarrow 0^-} p(t) = p_0 = \lim_{t\rightarrow 0^+} p(t), \ \ \ \ \lim_{t\rightarrow 0^-} q(t) = q_0 = \lim_{t\rightarrow 0^+} q(t)
\end{equation}
Suppose $\omega$ is increased from $\omega_0$ to $\omega_{f}$, the work is
\begin{eqnarray}
W_\tau(p_0,q_0) &=&  \lim_{t\rightarrow 0^+}H(p_0,q_0,t)-\lim_{t\rightarrow 0^-}H(p_0,q_0,t) \nn
                            &=& {m\over 2}(\omega_{f}^2-\omega_0^2)q_0^2.
\end{eqnarray}
The initial partition function $Z_0$ is still $2\pi /\beta \omega_0$, but this time we will express the distribution under $(p,q)$ coordinates,
\begin{equation}
\rho(p_0,q_0,0) ={1\over Z_0}\exp\left[{-\beta ({p_0^2\over 2m} +{m\omega_0^2\over 2}q_0^2)}\right].
\end{equation}
The work function is
\begin{eqnarray}
P(W)    =\int\limits_{\Gamma}{\beta \omega_0\over 2\pi}\exp\left[{-\beta ({p^2\over 2m} +{m\omega_0^2\over 2}q_0^2)}\right]\delta[W-{m\over 2}(\omega_{f}^2-\omega_0^2)q_0^2]dp_0dq_0 ,
\end{eqnarray}
noting there are $2$ distinct root of $W-{m\over 2}(\omega_{f}^2-\omega_0^2)q_0^2=0$,
\begin{eqnarray}
P(W)    &=&\int\limits_{\Gamma}{\beta \omega_0\over 2\pi}\exp\left[{-\beta ({p^2\over 2m} +{\omega_0^2\over \omega_{f}^2-\omega_0^2}W)}\right]2\times \biggr|{1\over 2\sqrt{W{m\over 2}(\omega_{f}^2-\omega_0^2)}}\biggr|dp_0 \nn
            &=&\int\limits_{\Gamma}{1\over\pi}\sqrt{\beta \omega_0^2 \over W(\omega_{f}^2-\omega_0^2) }\sqrt{\beta \over 2m}\exp\left[{-\beta ({p^2\over 2m} +{\omega_0^2\over \omega_{f}^2-\omega_0^2}W)}\right] dp_0 \nn
            &=&\sqrt{1\over\pi W} \sqrt{\beta \omega_0^2 \over (\omega_{f}^2-\omega_0^2) }\exp\left[-{\beta\omega_0^2\over \omega_{f}^2-\omega_0^2}W\right]
            \label{c4suddenwork}
\end{eqnarray}
for $W\ge 0$. The expectation of $W$ under sudden change is
\begin{eqnarray}
\<W\>_{sudden} &=& \int_0^\infty W \sqrt{1\over\pi W} \sqrt{\beta \omega_0^2 \over (\omega_{f}^2-\omega_0^2) }\exp\left[-{\beta\omega_0^2\over \omega_{f}^2-\omega_0^2}W\right] dW \nn
              &=& {1\over\sqrt\pi } {\omega_{f}^2-\omega_0^2  \over\beta \omega_0^2 }\Gamma({3\over 2}) \nn
              &=&{1\over 2} {\omega_{f}^2-\omega_0^2  \over\beta \omega_0^2 }.
              \label{c4suddenwork}
\end{eqnarray}
Here $\Gamma$ stand for Gamma function, not the phase space. Standard deviation of $W$ is calculated through
\begin{eqnarray}
\<W^2\>_{sudden} &=& \int_0^\infty W^2 \sqrt{1\over\pi W} \sqrt{\beta \omega_0^2 \over (\omega_{f}^2-\omega_0^2) }\exp\left[-{\beta\omega_0^2\over \omega_{f}^2-\omega_0^2}W\right] dW \nn
              &=& {1\over\sqrt\pi } \left({\omega_{f}^2-\omega_0^2  \over\beta \omega_0^2 }\right)^2\Gamma({5\over 2}) \nn
              &=& {3\over 4}\left({\omega_{f}^2-\omega_0^2  \over\beta \omega_0^2 }\right)^2,
\end{eqnarray}
and standard deviation $\sigma$ is
\begin{eqnarray}
\sigma_{sudden}(W) &=&\sqrt{ \<W^2\>_{sudden}- \<W\>_{sudden}^2}\nn
              &=& {1\over \sqrt 2}{\omega_{f}^2-\omega_0^2  \over\beta \omega_0^2 }.
\end{eqnarray}
And higher order moments can be easily obtained through Gamma function.

%%%%%%%%%%%%%%%%%%%%%%%%%%%%%%%%%%%%%%%%%
\section{Comparison Based on Numerical Results}
Although we have got the analytic solution for work function, it might be difficult to measure the work fluctuation for distributions like (\ref{c4nonadia}). Therefore we will basically use numerical results to compare different processes.

For this section, we will fix the parameter $m=1$ in our $H_0(p,q,t)$. Remember
\begin{equation}
\omega(t) =\omega_0 \sqrt{{f^2+1\over 2}-{f^2-1\over2}\cos (n\pi{t\over\tau})},
\end{equation}
we set $\omega_0 = 10$, $f=\sqrt 3$ and $n =1$. Here $\tau$ is an important parameter. It is the total duration of our process, therefore it together with $\omega$ describes adiabaticity of the process. Small $\tau\omega_0$ means the process tends to be non-adiabatic, while large $\tau\omega_0$ means adiabatic process.

Simulation follows the following steps. We first fix $\beta =1$, and then randomly take samples according to $\rho_0(p_0,q_0,0)$. Next we solve the corresponding differential equation numerically with initial condition $(p_0,q_0,0)$, and then calculate the work $W$. This procedure is repeated 1 million times, so that the histogram of $W$ will be a good estimation of work function $P(W)$. We simulate the following $3$ processes.

\

Process 1. Fast-Forward Adiabatic Process with Control Field $H_C$. We choose $\tau = 0.001$, such that $\tau\omega_0=0.01 \ll 2\pi$. This example is aimed to show we can keep the process adiabatic even if the parameter changes rapidly. The work function should be the same with the one of conventional adiabatic process

\

Process 2. Non-adiabatic Process without $H_C$. We choose $\tau = 0.001$, and $\tau\omega_0=0.01 $ in finite-time process. This example will show the work function of extremely fast and non-adiabatic process.

\

Process 3. Conventional Adiabatic Process without $H_C$. We choose $\tau = 10$, and $\tau\omega_0=100 $ in finite-time process. In this case, classical adiabatic theorem approximately holds, thus it could be regarded as conventional adiabatic process.

\subsection{Two Adiabatic Process}
We first compare the work function from process 1 and process 3. This is to convince the reader that fast-forward and conventional adiabatic process share the same work function.
\begin{figure}[here]
\centering
\includegraphics[scale=1.4]{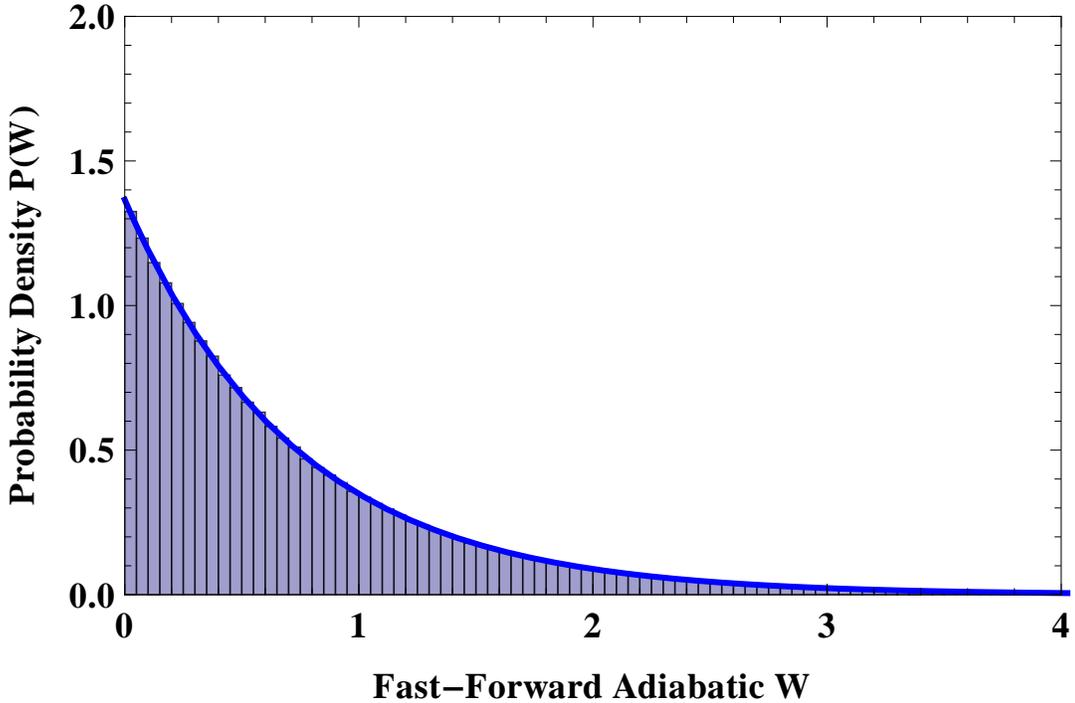}
\caption{Histogram of work function for fast-forward adiabatic process, $\tau\omega_0=0.001$, $\beta = 1$. Blue line is the theoretical work function given by(\ref{c21Drho}). }
\label{P(W)fastad}
\end{figure}
\begin{figure}[here]
\centering
\includegraphics[scale=1.4]{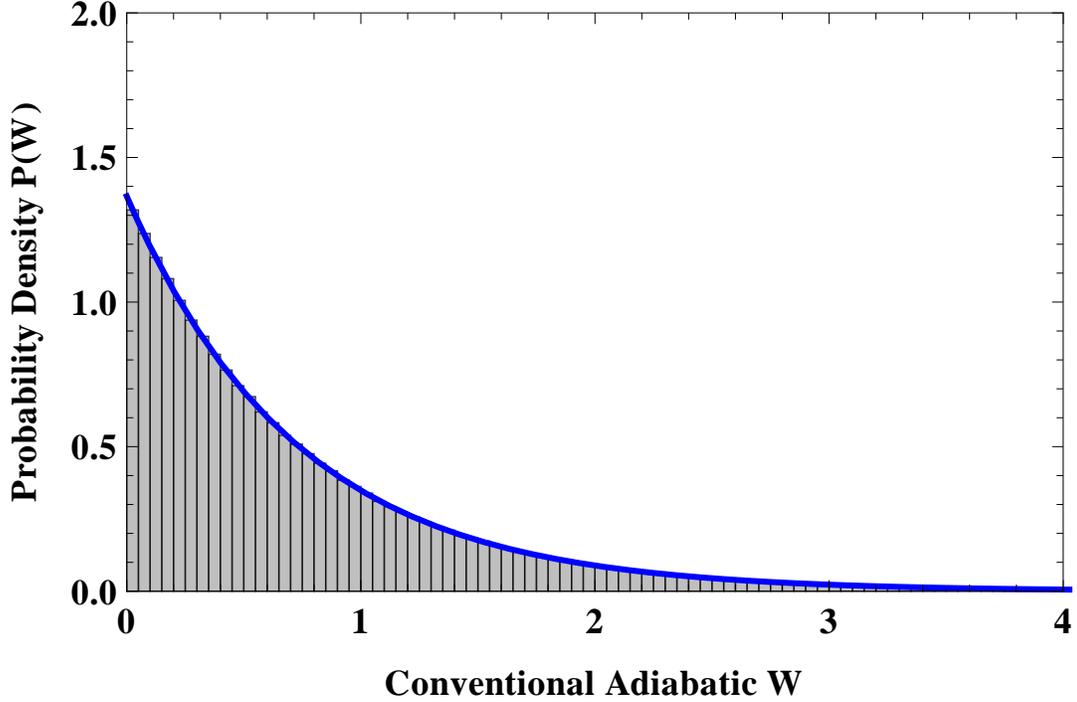}
\caption{Histogram work function for conventional adiabatic process, $\tau\omega_0=100$. Blue line is the theoretical work function given by (\ref{c4convenadw}).}
\label{P(W)ad}
\end{figure}

Figure \ref{P(W)fastad} and \ref{P(W)ad}  illustrate the work function of two adiabatic processes. The blue and gray bars are histogram of $W$ which estimates the work function. Blue lines indicate two identical theoretical work function (\ref{c21Drho}) and (\ref{c4convenadw}). We observe that these two distributions coincide with each other. Actually, we must split them into two graph in order to distinguish them from each other. Therefore we are quite confident that fast-forward adiabatic process has the identical work function with conventional adiabatic process.

\subsection{Fast-Forward Adiabatic versus Non-Adiabatic }
Here comes the crucial part. We are going to compare the work function of fast-forward adiabatic process (process 1) and non-adiabatic process (process 2) in this section. At the end of this section, we will reach our conclusion saying fast-forward adiabatic process has smaller work fluctuation, or equivalently, our control field $H_C$ significantly suppresses the work fluctuation.
\begin{figure}[here]
\centering
\includegraphics[scale=0.4]{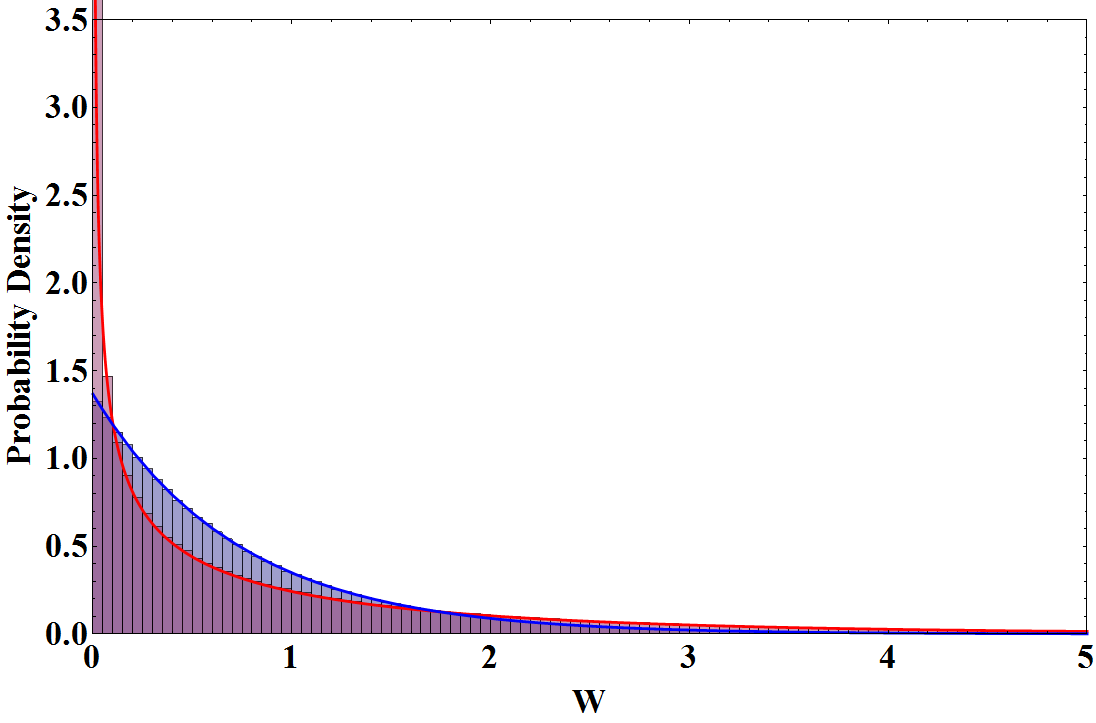}
\caption{Histogram and theoretically predicted curves of work function with $\beta = 1$. Blue one is for fast-forward adiabatic process (process 1), $\tau\omega_0=0.001$. Red one is for non-adiabatic process (process 2), $\tau\omega_0=0.001$. Theoretical predictions are given by (\ref{c21Drho}) and (\ref{c4nonadia}) respectively. Corresponding expectation and standard deviation are listed in Table \ref{table}. }
\label{P(W)}
\end{figure}

\begin{table}[htb]
\caption{List of expectation $\<W\>$ and standard deviation $\sigma(W)$ from simulation of fast-forward adiabatic process and non-adiabatic process. Theoretical predictions of $\<W\>$ and $\sigma(W)$ are also listed for comparison purpose.}
\begin{center}
\begin{tabular}{ccccc}
\hline
\hline
Process &   Simulated $\<W\>$   &Theoretical $\<W\>$   &Simulated $\sigma(W)$& Theoretical $\sigma(W)$
\\
\hline
$\begin{array}{c}
{\rm Fast-Forward}\\
{\rm Adiabatic}
\end{array}$ & 0.73192&$\sqrt 3-1 \approx 0.73205$ &0.73174&$\sqrt 3-1 \approx 0.73205$ \\
\hline
Non-Adiabatic & 0.99747&0.99999&1.40918&1.41419\\
\hline
\hline
\end{tabular}
\end{center}
\label{table}
\end{table}

\begin{figure}[h!]
\centering
\includegraphics[scale=0.5]{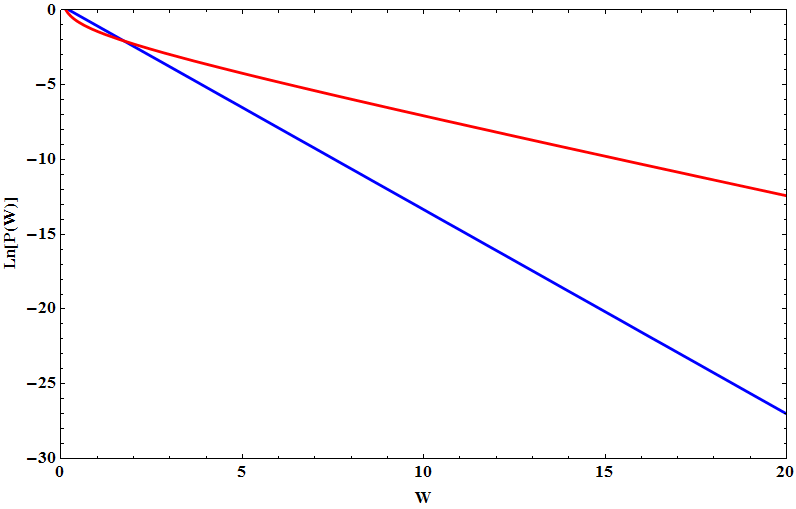}
\caption{Log plot of work function, i.e. $\ln P(W)$ versus $W$. Blue curve is for fast-forward adiabatic process (process 1). Red curve is for non-adiabatic process (process 2).}
\label{LogP(W)}
\end{figure}

After careful comparison, we find non-adiabatic process tends to spread out over $[0,\infty)$. According to Table \ref{table}, $\<W\>$ of non-adiabatic process is roughly around $W=1$, but non-adiabatic work function has a high peak at $W=0$, which is far from the mean value 1. This contributes to the large deviation of $W$. Moreover, fast-forward adiabatic work function converges to 0 much faster that non-adiabatic work function as $W\rightarrow \infty$. This feature is better illustrated in Figure \ref{LogP(W)}.

In Figure \ref{LogP(W)}, we can see as $W$ increases, $P(W)$ decreases much faster in fast-forward adiabatic case, which means non-adiabatic process will have a long tail when $W$ is large. The direct result is that the higher order moment, $\<W^n\>$ will increase faster in non-adiabatic process. So, not only the standard deviation $\sigma(W)$, but also higher order moments are larger in non-adiabatic process. Therefore we can conclude that $W$ fluctuates more violently in non-adiabatic process.

Noting that we choose the same parameters, including the duration of the process $\tau$ for these two processes. The only difference between them is the full Hamiltonian: fast-forward adiabatic process has an additional control field $H_C$. Thus another conclusion is that our control field $H_C$ could significantly suppress the work fluctuation.

Although our argument is based on 1-D harmonic oscillator, we believe our conclusion holds in general. This is because some important intermediate results is not restricted to harmonic oscillator case. For example, the work function of fast-forward adiabatic process should be identical with the one of conventional adiabatic process. Also, it is intuitive and reasonable to assume that non-adiabatic process has larger work fluctuation than adiabatic process. The above statements, combined with the fact that we can turn a non-adiabatic process to a fast-forward adiabatic one through control field, we can immediately summarize our control filed could suppress the work fluctuation.

%%%%%%%%%%%%%%%%%%
\chapter{Quantum Fast-Forward Adiabatic Process}
%%%%%%%%%%%%%%%%%%
Here comes our discussion on quantum fast-forward adiabatic process. The quantum version is actually based on Berry\rq{}s transitionless process \cite{Berry}. The idea is quite similar to our classical version. When ${\dot \lambda \rightarrow 0}$ is not satisfied, we will add a time-dependent control field $\hat H_C(t)$ onto the original Hamiltonian $\hat H_0(\lambda(t))$, such that the initial energy eigenstates of $\hat H_0(\lambda(0))$ will remain on the instantaneous eigenstate of $\hat H_0(\lambda(t))$. Explicit example of 1-D classical harmonic oscillator will be shown. Work function and work fluctuation will be compared for different processes.

\section{Control Field $H_C$}
In this section, we will use Berry\rq{}s approach. We first explore how the state evolves under adiabatic approximation, and then use these states to determine the time-evolution operator. The time-evolution operator will contain all the information about the full Hamiltonian $H$ in principle.

\subsection{Time-Evolution of Eigenstates}
Given a {\it non-degenerate} time-dependent Hamiltonian $\hat H_0(t)$, we could define the instantaneous energy eigenstates $|n(t)\>$ through
\begin{equation}
E_n(t)|n(t)\> = \hat H_0(t) |n(t)\>.
\label{eqn:staticstate}
\end{equation}
When ${\dot \lambda\over \lambda} \ll {\Delta E\over \hbar}$ is satisfied, the time evolution $|\psi_n(t)\>$ of initial state $|n(0)\>$ will remain on corresponding instantaneous eigenstate $|n(t)\>$, up to a phase factor, i.e.,
\begin{equation}
|\psi_n(t)\>\equiv \hat U(t,0)|n(0)\>\equiv \exp\left(-{i \over \hbar}\int_0^t dt\rq{}E_n(t\rq{})-\int_0^t dt\rq{}\<n(t\rq{})|\partial_{t\rq{}}n(t\rq{})\>\right) |n(t)\>,
\label{eqn:state}
\end{equation}
where $\hat U(t,0)$ is the time evolution, and the two integrals in exponential stand for dynamical and geometrical phase respectively.

\subsection{Formal Solution of $H_C$}
We are looking for $H_C(t)$ such that
\begin{equation}
{i} \hbar \partial_t|\psi_n(t)\> = \hat H(t)|\psi_n(t)\>,
\end{equation}
which implies
\begin{equation}
\hat H(t)\hat U(t)|n(0)\> =\hat  H(t)|\psi_n(t)\> = {i} \hbar \partial_t|\psi_n(t)\> =  {i} \hbar (\partial_t \hat U(t))|n(0)\>
\label{eqn:Shordinger}
\end{equation}
for all $|n(0)\>$. Or equivalently
\begin{equation}
\hat H(t) = {i} \hbar (\partial_t \hat U(t))\hat U^\dag(t),
\label{eqn:formalH}
\end{equation}
where $\hat H(t) =\hat H_0(t) + \hat H_C(t)$.

Notice that (\ref{eqn:state}) holds for any $|n(0)\>$, thus time evolution $\hat U(t,0)$ (or $\hat U(t)$ for short) satisfies
\begin{equation}
\hat U(t) = \sum\limits_{n} \exp\left(-{i \over \hbar}\int_0^t dt\rq{}E_n(t\rq{})-\int_0^t dt\rq{}\<n(t\rq{})|\partial_{t\rq{}}n(t\rq{})\>\right) |n(t)\>\<n(0)|,
\end{equation}
such that $|\psi_n(t)\> = \hat U(t)|n(0)\>$. Once we get $\hat U(t)$, by (\ref{eqn:Shordinger})
\begin{eqnarray}
\hat H(t) &=& {i} \hbar \partial_t (\hat U(t))\hat U^\dag(t) \nn
       &=& \sum\limits_{n} E_n |n(t)\>\<n(t)|+ {i} \hbar \sum\limits_{n}\left(|\partial_t n(t)\>\<n(t)|-\<n(t)|\partial_t n(t)\>|n(t)\>\<n(t)|\right).
\end{eqnarray}
Obviously the first summation is exactly $\hat H_0(t)$, thus the second summation is the $\hat H_C(t)$ we are looking for.

 Since we assume the system is non-degenerate, applying (\ref{c1compact}) will give us the compact form of $\hat H_C(t)$.
\begin{eqnarray}
\hat H_C(t) &=& {i} \hbar \sum\limits_{n}(|\partial_t n\>\<n|-\<n|\partial_t n\>|n\>\<n|) \nn
          &=&  {i} \hbar \sum\limits_{n} \sum\limits_{m} (|m\>\<m|\partial_t n\>\<n|-\<m|\partial_t n\>|m\>\<n|\delta_{mn})  \nn
          &=& {i} \hbar \sum\limits_{n} \sum\limits_{m \neq n} |m\>\<m|\partial_t n\>\<n| \nn
          &=&{i} \hbar \sum\limits_{n} \sum\limits_{m\neq n} {|m\>\<m|\partial_t \hat H_0(t) |n\>\<n| \over E_n - E_m},
          \label{eqn:quantumHC}
\end{eqnarray}
which is the expression worked out by Berry.

\subsection{Gibbs Canonical Ensemble}
The Gibbs canonical ensemble for quantum system is similar. Partition function $Z$ at time $t=0$ is
\begin{equation}
Z(0) = \sum\limits_{n} {e^{-\beta E_n(0)}},
\end{equation}
and initial mixed state is the same with (\ref{c2gibbs})
\begin{equation}
\rho(0) \equiv {1\over Z(0)}e^{-\beta \hat H_0(\lambda(0))} =  \sum\limits_{n} {e^{-\beta E_n(0)}\over Z(0)} |n(0)\>\<n(0)|.
\end{equation}
Corresponding work function is given by (\ref{c1quantumwf})
\begin{equation}
P(W) = \sum\limits_{n}\sum\limits_{m}P_n P_{n\rightarrow m}\delta \Bigl( W-[E_m(\tau)-E_n(0)]\Bigr),
\label{c5workfunction}
\end{equation}
where $P_n = {e^{-\beta E_n(0)}/ Z(0)}$, and $P_{n\rightarrow m}$ depend on specific process. Particularly, when the system undergoes conventional or fast-forward adiabatic process, there is no transition between states. In such cases, $P_{n\rightarrow m}=\delta_{mn}$, and
\begin{equation}
P(W) = \sum\limits_{n}P_n \delta \Bigl( W-[E_n(\tau)-E_n(0)]\Bigr).
\label{c5adwork}
\end{equation}
Notice this work function holds for any realization of parameter $\lambda(t)$, as well as the conventional adiabatic process.
%%%%%%%%%%%%%%%%%%%%%%%%%%%%%%
\section{1-D Quantum Harmonic Oscillator}

\subsection{Fast-Forward Adiabatic Process}
The original Hamiltonian is $\hat H_0(t) = {\hat p^2\over 2m} + {m\over 2} \omega^2(t)\hat q^2$. In the following calculation we will omit $t$ in $\omega$ and $\hat H$ for convenience.

To apply (\ref{eqn:quantumHC}) to our quantum harmonic oscillator, we first need to know
\begin{equation}
E_n-E_m = \hbar \omega (n-m) \neq 0,
\end{equation}
and
\begin{equation}
\partial_t \hat H_0= m \dot\omega \omega \hat q^2.
\label{SHO1}
\end{equation}
Notice that
\begin{eqnarray}
\hat a &=&\sqrt{m\omega \over 2\hbar} \left(\hat q + {{i} \over m \omega} \hat p \right),  \nn
\hat a^{\dag} &=&\sqrt{m \omega \over 2\hbar} \left(\hat q - {{i} \over m \omega} \hat p \right),
\end{eqnarray}
or
\begin{equation}
\hat q = \sqrt{ \hbar \over 2m\omega}(\hat a+\hat a^{\dag} ).
\label{SHO2}
\end{equation}
Plug into (\ref{SHO1}) and (\ref{SHO2}) into (\ref{eqn:quantumHC}), and we use $M$ for mass to distinguish it from quantum number $m$
\begin{eqnarray}
\hat H_C(t) &=& {i} \hbar \sum\limits_{n} \sum\limits_{m\neq n} {|m\>\<m|\partial_t \hat H_0(t) |n\>\<n| \over E_n - E_m} \nn
&=& {i} \hbar \sum\limits_{n} \sum\limits_{m\neq n} {|m\>\<m| M \dot\omega \omega  {\left[\sqrt{ \hbar \over 2M\omega}(\hat a+\hat a^{\dag} )\right]}^2 |n\>\<n| \over \hbar \omega (n-m)}  \nn
&=& {{i}\hbar \dot\omega \over 2\omega}\sum\limits_{n} \sum\limits_{m\neq n}{|m\>\<m|(\hat a^2+\hat a^{\dag 2} + \hat a \hat a^\dag +\hat a^\dag \hat a) |n\>\<n|\over  (n-m)} \nn
&=& {{i}\hbar \dot\omega \over 2\omega}\sum\limits_{n} \sum\limits_{m\neq n}{|m\> \<n|\over   (n-m)}(\sqrt{(m+1)(m+2)}\delta_{m+2,n}+(\sqrt{(n+1)(n+2)}\delta_{m,n+2} \nn
&\ & \ \ \ + \sqrt{(m+1)(n+1)}\delta_{m+1,n+1} + \sqrt{mn}\delta_{m-1,n-1}) \nn
(m\neq n)&=&{{i}\hbar \dot\omega \over 2\omega}\sum\limits_{n} {1\over 2 }(\sqrt{(n-1)n}|n-2\> \<n|-\sqrt{(n+1)(n+2)}|n+2\> \<n|) \nn
&=&{{i} \hbar\dot\omega \over 4\omega}\sum\limits_{n} (\hat a^2|n\> \<n|-\hat a^{\dag 2}|n\> \<n|) \nn
&=&{{i} \hbar\dot\omega \over 4\omega}(\hat a^2-\hat a^{\dag 2}) \nn
&=&{{i} \hbar\dot\omega \over 4\omega}{M\omega \over 2\hbar}\left[ {\left(\hat q + {{i} \over M \omega} \hat p \right)}^2-{\left(\hat q - {{i} \over M \omega} \hat p \right)}^2\right] \nn
&=& {{i}\hbar \dot\omega \over 4\omega}{M\omega \over 2\hbar}{2{i} \over M\omega}(\hat q\hat p + \hat p\hat q) \nn
&=& -{ \dot\omega \over 4\omega}(\hat q\hat p + \hat p\hat q),
\end{eqnarray}
which is consistent with our classical result $- \dot\omega pq / 2\omega$.

Now suppose we start with a canonical ensemble
\begin{equation}
\rho(0) \equiv {1\over Z(0)}e^{-\beta \hat H_0(\omega(0))} =  \sum\limits_{n=0}^\infty {e^{-\beta \hbar\omega_0 (n+{1\over 2})}\over Z(0)} |n(0)\>\<n(0)|,
\end{equation}
where $\omega_0\equiv \omega(0)$ and
\begin{equation}
Z(0) = \sum\limits_{n=0}^\infty {e^{-\beta\hbar\omega_0 (n+{1\over 2})}}={e^{-{1\over 2}\beta\hbar\omega_0 }\over 1-e^{-\beta\hbar\omega_0 }}.
\end{equation}
Under control field $\hat H_C$, there is not transition between states, thus the work function is
\begin{eqnarray}
P(W) &=& \sum\limits_{n=0}^\infty P_n \delta \Bigl( W-[E_n(\tau)-E_n(0)]\Bigr) \nn
         &=& \sum\limits_{n=0}^\infty {e^{-\beta \hbar\omega_0 (n+{1\over 2})}\over Z(0)}\delta \Bigl( W-[\hbar\omega_{f} (n+{1\over 2})-\hbar\omega_0 (n+{1\over 2})]\Bigr) \nn
         &=& \sum\limits_{n=0}^\infty  (1-e^{-\beta\hbar\omega_0 }) e^{-n\beta \hbar\omega_0 }\delta \Bigl( W-\hbar(\omega_{f}-\omega_0) (n+{1\over 2})\Bigr)
\label{c5adwork}
\end{eqnarray}

\subsection{Finite-Time Process}
In this section we will follow Lutz\rq{}s formalism \cite{Lutz}. Since his approach is quite complete for 1-D harmonic oscillator, we will only  introduce the brief steps.

According to Husimi \cite{Husimi}, given any time-dependent harmonic oscillator $\hat H_0(p,q,\omega(t)) = p^2/2m + m\omega^2(t) q^2$, the system is solvable. As if we can solve the corresponding classical Hamiltonian system $H_0(p,q,\omega(t))$ and get the specific solutions $S(t)$ and $C(t)$ as (\ref{S&C}) defines:
\begin{eqnarray}
C(0) = 1, \ \ &\ & \ \dot C(0)=0\ ; \nn
S(0) = 0, \ \ &\ &\ \dot S(0)=1.
\end{eqnarray}
The propagator in $x,x_0$ representation is given by
\begin{equation}
U(x,x_0;\tau)=\sqrt{m \over 2\pi {i} h S(\tau)}\exp\left[ {{i}m\over 2\hbar S(\tau)} (\dot S(\tau) x^2 -2xx_0 +C(\tau) x_0^2) \right]
\end{equation}
and final wavefunction
\begin{equation}
\phi(x,\tau) = \int dx_0 U(x,x_0;\tau)\phi(x_0,0)
\end{equation}
with initial wavefunction $\phi(x_0,0)$. Hence given any energy eigenfunction $n(x_0,0) = \<x_0|n(0)\>$ as initial state, in principle we could solve for the final state $\psi_n(x,\tau)$ using this propagator. And transition probability $P_{n\rightarrow m}$ can then be obtained. Although some calculation might not be doable analytically, we can always get a value through numerical integration.

Of course, this is not the end of the story. A compact form of work function can be obtained by introducing the characteristic function of work function $P(W)$. Characteristic function $G(\mu)$ is defined through Fourier transformation
\begin{equation}
G(\mu) \equiv \int dW e^{{i}\mu W}P(W).
\end{equation}
Characteristic function of work function is an alternative approach when $P(W)$ could not be solved exactly. Fortunately, $G(\mu)$ could be worked out under several extreme case, for example classical limit and adiabatic limit, thus we are able to compare and contrast the results under such cases.

%%%%%%%%%%%%%%%%%%%%%%%%%%%%%%
\section{Comparison Based on Numerical Results}
In order to compare the work properly, we will let control field $H_C = 0$ at both ends of the process,  by choosing $\omega(t)$ to be
\begin{equation}
\omega(t) =\omega_0 \sqrt{{f^2+1\over 2}-{f^2-1\over2}\cos (n\pi{t\over\tau})}
\label{c5omega}
\end{equation}
Quite similar to the classical case, most parameters are fixed: $\omega_0 = 10$, $f=\sqrt 3$, $n =1$ and mass $m=1$. We will further assume $\hbar = 1/2\pi$ in simulation.

\subsection{Simulation Results}
Since we have a quite simple work function (\ref{c5adwork}) for fast-forward adiabatic process , we only do the simulation of non-adiabatic process by choosing $\tau\omega_0 = 0.001$ without $\hat H_C$

According to (\ref{c5workfunction}) we only need to know $P_n$ and $P_{n\rightarrow m}$. We choose $\beta = 0.1$ and $P_n$ are directly calculated for Gibbs canonical ensemble.

\begin{figure}[here]
  \centering
    \includegraphics[scale=0.6]{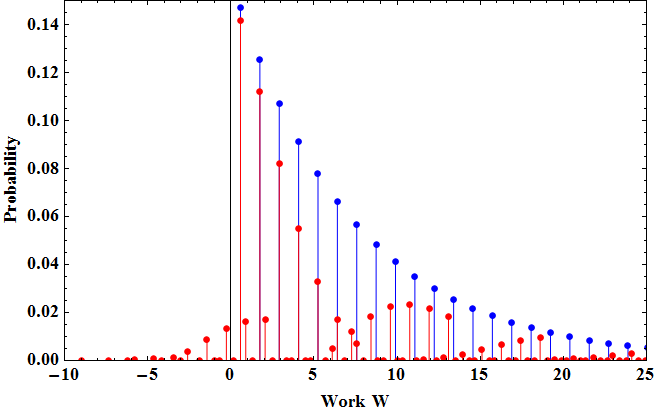}
    \caption{Red dots are simulation results for probability distribution of $W$ under non-adiabatic process $\tau\omega_0 = 0.001$ without control field. Blue dots represent theoretical distribution of fast-forward adiabatic process. Since the work function is summation over delta-function, height of the dots represents the probability instead of probability density. $\beta = 0.1$.}
    \label{fig:QuantumWorkSim}
\end{figure}

To get $P_{n\rightarrow m}$ we have to know how eigenstates evolves under $\hat H_0(t)$. Naturally we choose the eigenstate $|n(0)\>$ of $\hat H_0(0)$ as initial states. For each initial state, we simulate the time-evolution and find the corresponding wavefunction $\phi_n(x,\tau)$ at $t=\tau$ using split operator method. We then calculate the probability $P_{n\rightarrow m}$ of $\phi_n(x,\tau)$ falling into the instantaneous eigenstate $|m(\tau)\>$ of $\hat H_0(\tau)$. This process is repeated for $n=0$ to 7 and $m=0$ to 19. The probability distribution is show in Figure \ref{fig:QuantumWorkSim}.

From this graph, we first observe that the non-adiabatic work distribution spreads over real axis, but the adiabatic work distribution concentrates on several fixed value. More importantly, there exists negative work in non-adiabatic process.

\

The negative work is resulted from the quantum nature of this system. Since the process is non-adiabatic, it is possible for the state to jump from high-energy to low-energy, as Figure \ref{fig:QuantumWorkSim} indicates. And there are many different $W$ for a certain initial state. Note there is a unique $W$ in classical case if initial condition is given. The exact analog with classical work function needs $\<W\>_n$. Here $\<W\>_n$ stands for the expectation of work when we choose $|n(0)\>$ as our initial state. And this \lq\lq{}work function\rq\rq{} is given by
\begin{equation}
P\rq{}(W) = \sum\limits_{n}P_n \delta(W-\<W\>_n),
\end{equation}
where $P_n$ is the probability in Gibbs canonical ensemble
\begin{equation}
\rho = \sum\limits_{n}P_n |n\>\<n|.
\end{equation}
Using this approach, we could get rid of negative work in quantum non-adiabatic process.

\

Negative work is only one reason of the larger fluctuation in quantum non-adiabatic process. Another reason is again the long tail issue. All these features determine that non-adiabatic process has a larger work fluctuation.

\subsection{Classical Limit}
The quantum adiabatic work function is exactly the same with its classical correspondence. This result is obvious once we notice (\ref{c5adwork}) is a geometric distribution, which is a discrete analog of the exponential distribution. Preliminarily, we could show that the classical limit of quantum harmonic oscillator is congruent with the classical one, as Table \ref{table2} shows below.

\begin{table}[htb]
\caption{List of $\<W\>$, $\<W^2\>$ and standard deviation $\sigma(W)$ by theoretical prediction.}
\begin{center}
\begin{tabular}{cccc}
\hline
\hline
Process &    $\<W\>$   &$\<W^2\>$   &$\sigma(W)$
\\
\hline
Classical Adiabatic & $\sqrt 3-1 \approx 0.73205$&$2(\sqrt 3-1)^2 \approx 1.07180$ &$\sqrt 3-1 $\\
\hline
Classical non-Adiabatic & 0.99999&2.99993&1.41419\\
\hline
Quantum Adiabatic & $\sqrt 3-1 \approx 0.73205$&$2(\sqrt 3-1)^2 \approx 1.07180$ &$\sqrt 3-1 $\\
\hline
Quantum non-Adiabatic  & 1.00000& 3.00000&1.41421 \\
\hline
\hline
\end{tabular}
\end{center}
\label{table2}
\end{table}

%%%%%%%%%%%%%%%%%%
\chapter{Application}
%%%%%%%%%%%%%%%%%%
From previous chapters we know two outstanding features of our fast-forward adiabatic process and control field. First one is the control field greatly suppresses the work fluctuation. Second one is that it allows adiabatic process to be performed arbitrarily fast, which saves much time in practice. In this chapter we are going to discuss how these two features are applied in Jarzynski equality and quantum engine.

\section{Revisit Jarzynski Equality with Control Field}
\begin{figure}[here]
  \centering
    \includegraphics[scale=0.65]{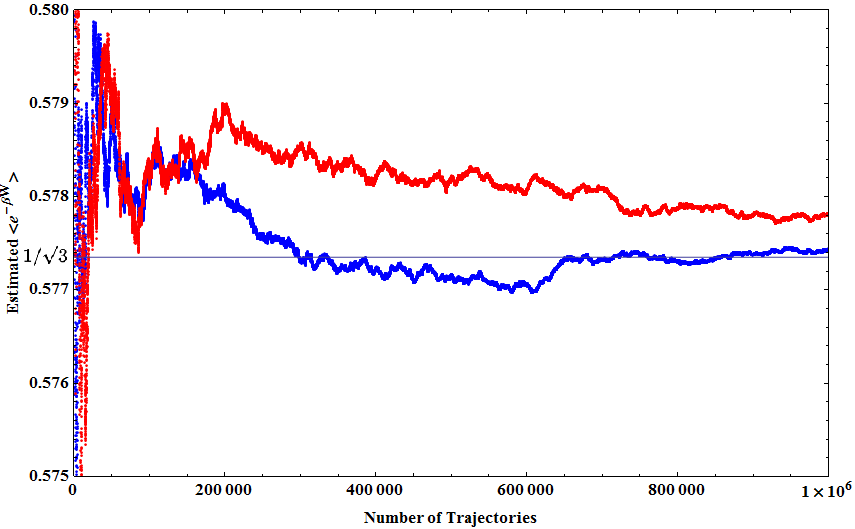}
    \caption{Simulation results of converge tendency of fast-forward adiabatic process and non-adiabatic process. $\omega_0 = 10$,$\beta =1$. $\omega_0 \tau= 0.001$. Red curve is for non-adiabatic process, blue one for fast-forward adiabatic process. The middle horizontal line is the theoretical value of $\<e^{-\beta W}\>$, $1/\sqrt3$.}
    \label{fig:convergingrate}
\end{figure}

From chapter three we know that Jarzynski relates the non-equilibrium quantity work $W$ with the equilibrium quantity free energy $F$ through
\begin{equation}
\<e^{-\beta W}\> = e^{-\beta \Delta F}.
\label{c6Jar}
\end{equation}

In chapter three we also discussed that Jarzynski equality contains only information about the average, and does not contain information about the work fluctuation. So if the work fluctuation is very large, converging rate of $\<e^{-\beta W}\>$ might be very slow. In such cases, if we add the control field $H_C$, the non-equilibrium process can be completed with smaller work fluctuation and within the same time. Figure \ref{fig:convergingrate} shows the comparison between simulation results of non-adiabatic process without control field or fast-forward adiabatic process.

\

From this figure, we find that the estimated $\<e^{-\beta W}\>$ of fast-forward adiabatic process approached theoretical value much faster. It comes stable around the theoretical value $1/\sqrt 3$ after $3\times 10^5$ trajectories. However, the non-adiabatic one comes to $1/\sqrt 3$ at the end of simulation --- $1\times 10^6$ trajectories. Hence $H_C$ could speed up converging of Jarzyski average $\<e^{-\beta W}\>$ by reducing the work fluctuation.

%%%%%%%%%%%%%%%%%%%%
\section{Quantum Engine}
\subsection{Quantum Otto Cycle}
\begin{figure}[here]
  \centering
    \includegraphics[scale=0.5]{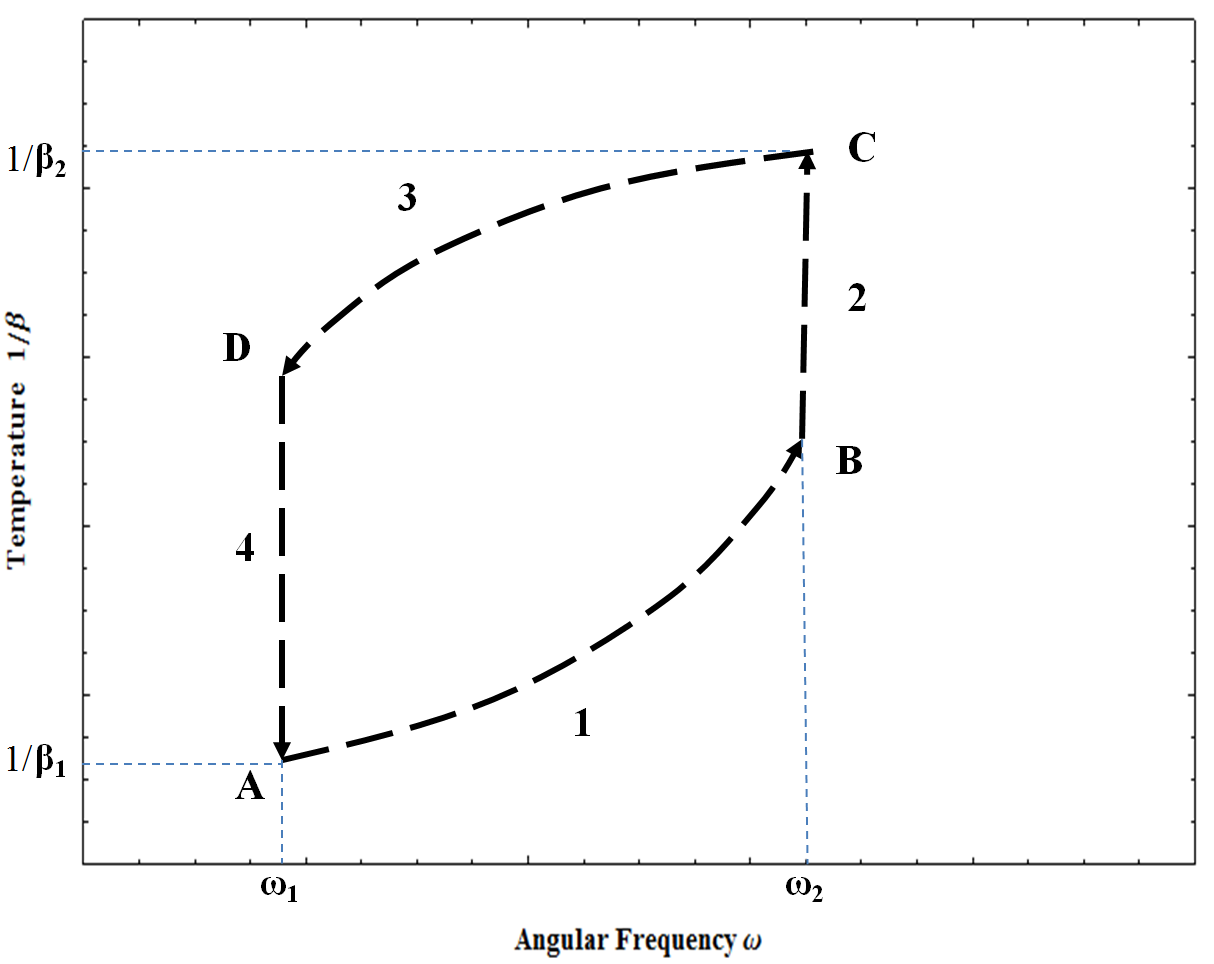}
    \caption{Sketch map of Otto cycle}
    \label{fig:otto}
\end{figure}

The quantum engine we are going to discuss is based on Otto cycle considered by Lutz \cite{LutzEng}. We suppose the system is a previously discussed quantum harmonic oscillator. The cycle consists of four consecutive steps as shown in \ref{fig:otto}. Suppose the start point $A(\omega_1,\beta_1)$ is a Gibbs canonical ensemble.

\

1. Isentropic compression $A(\omega_1,\beta_1) \rightarrow B(\omega_2,*)$. The angular frequency is increased during time $\tau_1$, and the system is isolated from any heat reservoir. The time-evolution is unitary, thus the von Neumann entropy is constant. Notice $B$ is no longer an equilibrium state, thus we use $*$ instead of exact temperature.

2. Hot isochore $ B(\omega_2,*)\rightarrow C(\omega_2,\beta_2) $. The angular frequency of the system is fixed, and meanwhile the system is weakly coupled with a heat reservoir at $\beta_2$. Thus state $C$ is a canonical ensemble. The relaxation time is $\tau_2$.

3. Isentropic expansion $  C(\omega_2,\beta_2)\rightarrow D(\omega_1,*)$. $\omega$ is decreased to $\omega_1$ during time $\tau_3$ while system is isolated from heat reservoir.

4. Cold isochore $ D(\omega_1,*)\rightarrow A(\omega_1,\beta_1) $. Similar to 2. Time duration is $\tau_4$.

\

The authors consider two cases under classical limit: angular frequency in 1 and 3  is changed slowly (conventional adiabatic limit) or fast (sudden change limit), and then consider the efficiency at maximum average output for a cycle. Here, for convenience we assume the oscillator is actually a classical one. There are two reasons. First, the results of both classical oscillator and quantum oscillator under classical limit turn out to be the same. It is not surprising that classical and quantum process share the same work function under classical limit $\hbar\beta \ll 1$. Second, many bio-motors are actually classical engines as they are under room temperature, which will kill most quantum effects. Thus it is reasonable to apply the results of classical fast-forward adiabatic process to quantum engines under classical limit.

\

We first calculate the adiabatic limit case. Average energy of state $\<E_A\>$ is $1/\beta_1$, which is obvious. And by (\ref{c21Drho}), the average work done in 1 is $\<W_1\> = \Delta \omega /\beta_1 \omega_1 = (\omega_2-\omega_1)/\beta_1 \omega_1$. Thus average energy of state $B$ is $\<E_B\>=\<W_1\>+\<E_A\>=\omega_2/\beta_1 \omega_1$. Since $\<E_C\> = 1/\beta_2$, the heat received from hight-temperature reservoir is $\<Q_2\> = \<E_C\>-\<E_B\> = 1/\beta_2-\omega_2/\beta_1 \omega_1$. We could get $\<W_3\> = (\omega_1-\omega_2)/\beta_2 \omega_2 $ in the similar way. Thus the average work done in one cycle is
\begin{equation}
\<W_1\>+\<W_3\> = {1\over \beta_1}{\omega_2-\omega_1\over \omega_1}+{1\over \beta_2}{\omega_1-\omega_2\over \omega_2}.
\end{equation}
Noting the total time is $\tau_1+\tau_2+\tau_3+\tau_4$, and output $P = -(\<W_1\>+\<W_3\> )/(\tau_1+\tau_2+\tau_3+\tau_4)$. The tricky part is, $\tau_1+\tau_3$ are dominant, because 1 and 3 are adiabatic process. Therefore minimizing total work in one cycle is actually maximizing the output. After simple mathematics, we find the maximum power occurs at $\omega_2/\omega_1 = \sqrt{\beta_1/\beta_2}$. The efficiency at maximum output is given by
\begin{eqnarray}
\eta_{ad} &=&- {\<W_1\>+\<W_3\>\over \<Q_2\>} \nn
    &=& 1-\sqrt{\beta_2\over \beta_1}.
\end{eqnarray}
The power adiabatic limit is very low since $\tau_1+\tau_3$ tends to infinity.

The sudden limit case is quite similar. Using (\ref{c4suddenwork}) $\<W_1\> ={(\omega_{2}^2-\omega_1^2 ) /2\beta_1 \omega_1^2 }$, and $\<E_B\>={(\omega_{2}^2+\omega_1^2 ) /2\beta_1 \omega_1^2 }$. Then $\<Q_2\> = \<E_C\>-\<E_B\> = 1/\beta_2-(\omega_{2}^2+\omega_1^2 ) /2\beta_1 \omega_1^2 $, and $\<W_3\> ={(\omega_{1}^2-\omega_2^2 ) /2\beta_2 \omega_2^2 }$. This time, under sudden change limit, total time is $\tau_2+\tau_4$ as $\tau_1,\tau_3\rightarrow 0$. We further assume $\tau_2+\tau_4$ is approximately constant for any combination of $\omega_1, \omega_2$. This is reasonable as they are system relaxation time. So minimizing total work
\begin{equation}
\<W_1\>+\<W_3\> = {(\omega_{2}^2-\omega_1^2 ) \over 2\beta_1 \omega_1^2 }+{(\omega_{1}^2-\omega_2^2 ) \over 2\beta_2 \omega_2^2 }
\end{equation}
gives us $\omega_2/\omega_1 = \sqrt[4]{\beta_1/\beta_2}$. And efficiency at maximum output is
\begin{eqnarray}
\eta_{su} &=& -{\<W_1\>+\<W_3\>\over \<Q_2\>} \nn
    &=&{ 1-\sqrt{\beta_2/\beta_1}\over 2+\sqrt{\beta_2/\beta_1}} <{1\over 2}\eta_{ad}.
\end{eqnarray}
The power of sudden change is much bigger than adiabatic limit, but the efficiency is less than half of $\eta_{ad}$.

\subsection{Quantum Engine with $H_C$}
In previous section we show that designers of quantum engine have to choose between efficiency and work output: sudden change provides higher output while conventional adiabatic process provides higher efficiency. Now let\rq{}s apply our fast-forward adiabatic process to 1 and 3. Obviously the efficiency at maximum output should be the same for both conventional and fast-forward adiabatic process, since they share the same work function, expectation and fluctuation, etc. But the output can be improved up to $- (\<W_1\>+\<W_3\>)/(\tau_2+\tau_4)$ because the fast-forward process can be performed arbitrarily fast.

Next compare it with the sudden change limit. We know the efficiency is improved more than twice, how about the output power? For fast-forward adiabatic process, maximum power occurs at $\omega_2/\omega_1 = \sqrt{\beta_1/\beta_2}$, the absolute value of total work is
\begin{eqnarray}
|\<W_1\>+\<W_3\> |_{ad}&=& {1\over \beta_1}{\omega_2-\omega_1\over \omega_1}+{1\over \beta_2}{\omega_1-\omega_2\over \omega_2} \nn
   &=& \biggr|{1\over \beta_1}\sqrt{\beta_1\over\beta_2}+{1\over \beta_2}\sqrt{\beta_2\over\beta_1}-{1\over \beta_1}-{1\over \beta_2}\biggr|\nn
   &=&\biggr|{2\over \sqrt{\beta_1\beta_2}}-{1\over \beta_1}-{1\over \beta_2}\biggr|
\end{eqnarray}
For sudden change, maximum power is at $\omega_2/\omega_1 = \sqrt[4]{\beta_1/\beta_2}$,
\begin{eqnarray}
|\<W_1\>+\<W_3\> |_{ad}&=& {(\omega_{2}^2-\omega_1^2 ) \over 2\beta_1 \omega_1^2 }+{(\omega_{1}^2-\omega_2^2 ) \over 2\beta_2 \omega_2^2 } \nn
   &=& \biggr|{1\over 2\beta_1}\sqrt{\beta_1\over\beta_2}+{1\over 2\beta_2}\sqrt{\beta_2\over\beta_1}-{1\over 2\beta_1}-{1\over 2\beta_2}\biggr|\nn
   &=&{1\over 2}\biggr|{2\over \sqrt{\beta_1\beta_2}}-{1\over \beta_1}-{1\over \beta_2}\biggr| \nn
   &=&{1\over 2}|\<W_1\>+\<W_3\> |_{ad},
\end{eqnarray}
which means fast-forward adiabatic process doubles both the efficiency and output if we assume $\tau_2+\tau_4$ is constant.

Besides the output and efficiency issue, there are other advantages of fast-forward adiabatic process. Quantum engine, as the name suggests, is engine work on small scale system. Due to the size of the system, the fluctuation is not negligible. Large work fluctuation, especially negative work in quantum non-adiabatic work function, might lead to a fluctuated output. However, quite uniform output of a heat engine is always one important industrial requirement, and fast-forward adiabatic process suppresses the fluctuation, i.e. provides a much more uniform output.

%%%%%%%%%%%%%%%%%%
\chapter{Conclusion}
%%%%%%%%%%%%%%%%%%

In this report, we discuss the fast-forward adiabatic process, particularly its effect in suppressing the work fluctuation in details.

We review both the classical and quantum adiabatic theorems, which describe the slowly changing feature of the conventional adiabatic process. We emphasize and rigorously define the work function of small system, as well as the work fluctuation.

In classical aspect, we construct our original control field under the most general condition, which could turn a non-adiabatic process into a fast-forward process. We calculate an explicit example in a time-dependent harmonic oscillator to illustrate how to construct the control field analytically. Numerical and simulated results are performed in order to compare the work functions of different processes.

In quantum aspect, we follow the works of Berry and Lutz on transitionless driving and non-adiabatic process. We make our contribution by comparing them. We propose to use control field to make a fast non-adiabatic process adiabatic, which effectively suppresses the work fluctuation.

We verify our arguments again using a time-dependent harmonic oscillator example. The toy model also reveals physical intrinsics, for example, how the quantum nature affects the work function. Based on our formalism and examples in fast-forward adiabatic process, we conjecture that the work fluctuation argument holds in general, which is resulted from the nature of adiabatic assumption.

There are many applications of the fast-forward adiabatic processes. In this report we only briefly touch two of them. The first one is the Jarzynski equality which links the thermo-average of the work function and the change of free energy. Using fast-forward adiabatic processes, the equality converges much faster. The second application is a quantum engine based on an Otto cycle. The application of fast-forward adiabatic processes not only maximizes the power output by speeding up the Otto cycle, but also increases the efficiency of the engine.

The results in this report can be easily realized in models other than the one-dimensional harmonic oscillator, for example, the two-level system. The fast-forward adiabatic processes can be also applied to quantum engines based on other cycles.

%%%%%%%%%%%%%%%%%%%%%%%%%%%%%%
%\begin{appendices}
% \renewcommand{\thechapter}{\Alph{chapter}.}
%
%
%\end{appendices}

%%%%%%%%%%%%%%%%%%%%%%%%%%%%%%%%%%%


\begin{thebibliography}{}
\bibitem[1]{Huang}K. Huang, {\it Statistical Mechanics} (Wiley, 1987), 2nd ed.
\bibitem[2]{JarFluc}C. Jarzynski and D. K. Wojcik, Phys. Rev. Lett. 92,230602 (2004).
\bibitem[3]{HanReview}M. Campisi, P. Hanggi, and P. Talkner, Rev. Mod. Phys. 83,771 (2011).
\bibitem[4]{Jar}C. Jarzynsk,  Phys. Rev. Lett. 78.2690(1997).
\bibitem[5]{HanOpen}M. Campisi, P. Talkner, and P. Hanggi, Phys. Rev. Lett. 102, 210401 (2009).
\bibitem[6]{Lutz}S. Deffner and E. Lutz, Phys. Rev. E 77, 021128 (2008).
\bibitem[7]{Berry}M. V. Berry, J. Phys. A: Math. Theor. 42 365303 (2009).
\bibitem[8]{LutzEng}O.Abah, J. Rossnagel, G. Jacob, S. Deffner, F. Schmidt-Kaler, K. Singer, and E, Lutz, Phys. Rev. Lett. 109, 203006 (2012).
\bibitem[9]{Chen2}Xi Chen, I. Lizuain, A. Ruschhaupt, D. Guéry-Odelin, and J. G. Muga, Phys. Rev. Lett. 105, 123003 (2010).
\bibitem[10]{ChenTran}Xi Chen, E. Torrontegui, D. Stefanatos, Jr-Shin Li, and J. G. Muga, Phys. Rev. A 84, 043415 (2011).
\bibitem[11]{Bouch}M. A. Bouchiat and C. Bouchiat, Phys. Rev. A 83, 052126 (2011).
\bibitem[12]{Gong}Jiangbin Gong, Lecture Notes of {\it Advanced Dynamics}.
\bibitem[13]{Gold}H. Goldstein, C. Poole, and J. Safko, {\it Classical Mechanics} (Pearson, 2002), 3rd ed.
\bibitem[14]{Husimi}K. Husimi, Prog. Theo. Phys. 9, 381 (1953).

\end{thebibliography}
\end{document}